\documentclass[12pt]{article}
\usepackage{amsmath,amssymb,theorem,cite,epsfig,url,psfrag,eepic,mathtools}
\advance \topmargin by -\headheight
\advance \topmargin by -\headsep
\evensidemargin \oddsidemargin
\def\Maketitle{{\def\newpage{}\maketitle}}
\makeatletter
\def\Appendix{\appendix
  \def\@seccntformat##1{Appendix~\csname the##1\endcsname.~~}}
\makeatother
\makeatletter
\@addtoreset{equation}{section}

\makeatother
\theorembodyfont{\sffamily}
\newtheorem{theorem}{Theorem}[section]

\newtheorem{Proposition}[theorem]{Proposition}

\def\jac{\mathop{\bf J}\nolimits}

\def\XXint#1#2#3{{\setbox0=\hbox{$#1{#2#3}{\int}$}
\vcenter{\hbox{$#2#3$}}\kern-.5\wd0}}

\def\ñth{{\textrm{\,ñth}}}
\def \eps {\epsilon}
\begin{document}
\title{\textbf{Instanton moduli spaces and bases in  coset conformal field theory}\vspace*{.3cm}}
\author{A.~A.~Belavin$^{1}$, M.~A.~Bershtein$^{1,2}$, B.~L.~Feigin$^{1,2,3}$,\\ A.~V.~Litvinov$^{1,4}$ and G.~M.~Tarnopolsky$^{1,4}$\vspace*{10pt}\\[\medskipamount]
$^1$~\parbox[t]{0.88\textwidth}{\normalsize\it\raggedright
Landau Institute for Theoretical Physics,
142432 Chernogolovka, Russia}\vspace*{4pt}\\[\medskipamount]
$^2$~\parbox[t]{0.88\textwidth}{\normalsize\it\raggedright
Independent University of Moscow,
Russia, Moscow, 119002}\vspace*{4pt}\\[\medskipamount]
$^3$~\parbox[t]{0.88\textwidth}{\normalsize\it\raggedright
Higher School of Economics,
Russia, Moscow, 101000}\vspace*{4pt}\\[\medskipamount]
\hspace*{-7pt}
$^4$~\parbox[t]{0.88\textwidth}{\normalsize\it\raggedright
Kavli Institute for Theoretical Physics, University of California,
 Santa Barbara, CA 93106}}
\date{}
\rightline{NSF-KITP-11-213}
\Maketitle
\begin{abstract}\vspace*{2pt}
Recently proposed relation between conformal field theories in two dimensions and supersymmetric gauge theories in four dimensions predicts the existence of the distinguished  basis in the space of local fields in CFT. This basis  has a number of remarkable properties,  one of them is the complete factorization of the coefficients of the operator product expansion.  We consider a particular case of the $U(r)$ gauge theory on $\mathbb{C}^{2}/\mathbb{Z}_{p}$ which corresponds to a certain coset conformal field theory and describe the  properties of this basis. We argue that in the case $p=2$, $r=2$ there exist different bases. We give an explicit construction of one of them. For another basis we propose the formula for matrix elements.
\end{abstract}
\vspace*{0pt}
\section{Introduction}
Two-dimensional conformal field theories and  $\mathcal{N}=2$ supersymmetric gauge theories in four dimensions were developed independently through years. However, it was observed   in the paper by Alday, Gaiotto and Tachikawa \cite{Alday:2009aq}  that the instanton part of the partition function in  $\mathcal{N}=2$ gauge theory coincides with the conformal block in 2d conformal field theory.

The relation between these two different  types of theories  is carried out through the intermediate object --- moduli space of instantons $\mathcal{M}$:
\begin{equation}\label{scheme}
\begin{picture}(30,75)(160,10)
    \Thicklines
    \unitlength 2.3pt
    \put(20,15){\vector(1,1){10}}
    \put(20,15){\vector(-1,-1){10}}
    \put(120,15){\vector(1,-1){10}}
    \put(120,15){\vector(-1,1){10}}
    \put(34,28){\mbox{\large{Instanton moduli space $\mathcal{M}$}}}
    \put(-2,-5){\mbox{\large{CFT}}}
    \put(105,-5){\mbox{\large{$\mathcal{N}=2$ gauge theory}}}
    \end{picture}
\vspace*{1cm}
\end{equation}
The right arrow on this picture symbolises that the path integral for the partition function in $\mathcal{N}=2$ supersymmetric gauge theory  is localized  and can be reduced  to  the integral over the manifold $\mathcal{M}$  (manifold $\mathcal{M}$ is disconnected, its connected components are labeled by some topological characteristics of instantons). The last integral is divergent due to the non-compactness of  the manifold $\mathcal{M}$. However, one can introduce proper regularization in the gauge theory \cite{Nekrasov:2002qd} which breaks Lorenzian symmetry, but preserves some of the supersymmetries and makes it possible to apply the localization technique. The regularized  integral is localized at the  fixed points of an abelian group (torus) which acts on $\mathcal{M}$ by the space-time rotations survived after breaking of Lorenzian symmetry and by the gauge transformation at infinity. The advantage of using of the deformed theory is that the fixed points of the torus are isolated. Hence the partition function is given by the sum of the fixed points contributions. The partition function defined in such a way is usually referred as  Nekrasov partition function.

The non-trivial part of \eqref{scheme} is represented by the left arrow which means, that there is a natural action of the symmetry algebra $\mathcal{A}$ of some conformal field theory on equivariant cohomologies of $\mathcal{M}$ (see Nakajima's papers \cite{Nakajima_1995,0970.17017} for basic examples of such action). Basis in the (localized) equivariant  cohomology space can be labeled by the fixed points of the torus \cite{Atiyah_Bott_1984}. Thus the geometrical construction gives some special basis of states in the highest weight representations $\pi_{\mathcal{A}}$ of the algebra $\mathcal{A}$. This basis is already remarkable just because of its geometrical origin and possesses many nice properties. Let us list some of them:
\begin{itemize}
\item To every torus fixed point $p\in\mathcal{M}$ correspond basic vector $v_p \in \pi_{\mathcal{A}}$. Moreover if $p \in\mathcal{M}_N$, where $N$ is a topological number then the vector $v_p$ has degree $N$.
\item There is a geometrically constructed scalar product on $\pi_{\mathcal{A}}$. Basis $v_p$ is orthogonal under this product and the norm of the vector $v_p$ equals to the determinant of the vector field $v$ in the tangent space of $p$. The last expression is also denoted by $Z^{-1}_{\textsf{vec}}$ (contribution of the vector multiplet).
\item Matrix elements of geometrically defined vertex operators have completely factorized form. The last expressions are also denoted by $Z_{\textsf{bif}}$ (contribution of the bifundamental multiplet).
\item There is a commutative algebra (Integrals of Motion) which is diagonalized in the basis $v_p$. Geometrically this algebra arise from the multiplication on cohomology classes.
\end{itemize}
Knowledge of the functions  $Z_{\textsf{vec}}$ and $Z_{\textsf{bif}}$  allows to compute multi-point conformal blocks  on a surface of genus $0$ and $1$. In CFT they give explicit and remarkably simple expressions for the coefficients of the operator product expansion.

In this paper we consider the particular case of the scheme described above.  Namely, we consider the case when  $\mathcal{M}$ is the moduli space of $U(r)$ instantons on $\mathbb{C}^2/\mathbb{Z}_p$ where $\mathbb{Z}_p$ acts by formula ($z_{1}$ and $z_{2}$ are coordinates on $\mathbb{C}^{2}$)
$$
(z_1,z_2) \mapsto (\omega z_1,\omega^{-1}z_2), \quad \text{where}\quad
\omega^p=1.
$$
There are several smooth partial compactifications of this space. One of them can be constructed as follows. Denote by $\mathcal{M}(r,N)$ smooth compactified moduli space of $U(r)$ instantons on $\mathbb{C}^2$ with topological number $N$. The set $\mathcal{M}(r,N)^{\mathbb{Z}_{p}}$ of $\mathbb{Z}_p$-invariant  points on $\mathcal{M}(r,N)$ is a smooth compactification of the space of instantons on $\mathbb{C}^2/\mathbb{Z}_p$. The torus action on $\mathcal{M}(r,N)^{\mathbb{Z}_{p}}$ induced by the actions on $\mathbb{C}^{2}$ and on framing at infinity. The fixed points of this torus are labeled by $r$-tuples $(Y_{1},\dots,Y_{r})$ of Young diagrams colored in $p$ colors. Then, there should be  a basis labeled by $(Y_{1},\dots, Y_{r})$ in a representation of some algebra $\mathcal{A}$.

It was suggested in \cite{Belavin:2011pp} that the instanton manifold $\mathcal{M}=\bigsqcup_{N}\mathcal{M}(r,N)^{\mathbb{Z}_{p}}$ corresponds  to the  coset conformal field theory
\begin{equation}\label{BF-coset}
\mathcal{A}(r,p)\overset{\text{def}}{=}\frac{\widehat{\mathfrak{gl}}(n)_r}{\widehat{\mathfrak{gl}}(n-p)_r},
\end{equation}
where parameter $n$ is related to equivariant parameters and in general can be arbitrary complex number.
Using well known level-rank duality this coset can be rewritten as
\begin{equation}
\mathcal{A}(r,p)=
\widehat{\mathfrak{gl}}(p)_r\times \frac{\widehat{\mathfrak{gl}}(n)_r}{\widehat{\mathfrak{gl}}(p)_r \times \widehat{\mathfrak{gl}}(n-p)_r}=
\mathcal{H}
\times \widehat{\mathfrak{sl}}(p)_r\times \frac{\widehat{\mathfrak{sl}}(r)_p \times \widehat{\mathfrak{sl}}(r)_{n-p}}{\widehat{\mathfrak{sl}}(r)_n},
\end{equation}
where $\mathcal{H}$ is the Heisenberg algebra.
Taking into account the construction of \cite{Goddard:1986ee} some of these algebras can be rewritten as
\begin{subequations}\label{scheme-12}
\begin{equation}\label{scheme-1}
 \begin{picture}(300,80)(0,40)
    \Thicklines
    \unitlength 1.6pt
    \put(0,0){\line(1,0){200}}
    \put(0,20){\line(1,0){200}}
    \put(0,40){\line(1,0){200}}
    \put(0,60){\line(1,0){200}}
    \put(0,0){\line(0,1){75}}
    \put(60,0){\line(0,1){75}}
    \put(120,0){\line(0,1){75}}
    \put(180,0){\line(0,1){75}}
    \put(27,8){\mbox{\small{$\mathcal{H}$}}}
    \put(77,8){\mbox{\small{$\mathcal{H}\oplus\mathsf{Vir}$}}}
    \put(137,8){\mbox{\small{$\mathcal{H}\oplus\mathsf{W}_{3}$}}}
    \put(15,28){\mbox{\small{$\mathcal{H} \oplus \widehat{\mathfrak{sl}}(2)_1$}}}
    \put(15,48){\mbox{\small{$\mathcal{H} \oplus \widehat{\mathfrak{sl}}(3)_1$}}}
    \put(64,28){\mbox{\small{$\mathcal{H} \oplus \widehat{\mathfrak{sl}}(2)_2 \oplus \textsf{NSR}$}}}
    \put(-22,8){\mbox{\small{$p=1$}}}
    \put(-22,28){\mbox{\small{$p=2$}}}
    \put(-22,48){\mbox{\small{$p=3$}}}
    \put(22,-8){\mbox{\small{$r=1$}}}
    \put(82,-8){\mbox{\small{$r=2$}}}
    \put(142,-8){\mbox{\small{$r=3$}}}
    \put(25,70){\circle*{1}}
    \put(30,70){\circle*{1}}
    \put(35,70){\circle*{1}}
    \put(85,50){\circle*{1}}
    \put(90,50){\circle*{1}}
    \put(95,50){\circle*{1}}
    \put(85,70){\circle*{1}}
    \put(90,70){\circle*{1}}
    \put(95,70){\circle*{1}}
    \put(145,30){\circle*{1}}
    \put(150,30){\circle*{1}}
    \put(155,30){\circle*{1}}
    \put(145,50){\circle*{1}}
    \put(150,50){\circle*{1}}
    \put(155,50){\circle*{1}}
    \put(145,70){\circle*{1}}
    \put(150,70){\circle*{1}}
    \put(155,70){\circle*{1}}
    \put(195,50){\circle*{1}}
    \put(200,50){\circle*{1}}
    \put(205,50){\circle*{1}}
    \put(195,30){\circle*{1}}
    \put(200,30){\circle*{1}}
    \put(205,30){\circle*{1}}
    \put(195,10){\circle*{1}}
    \put(200,10){\circle*{1}}
    \put(205,10){\circle*{1}}
    \end{picture}
\vspace*{1.9cm}
\end{equation}
where \textsf{Vir} is the Virasoro algebra, $\mathsf{W}_{3}$ is the $\mathfrak{sl}(3)$ $W$ algebra and \textsf{NSR} is the Neveu--Schwarz--Ramond algebra, $N=1$ superanalogue of the Virasoro algebra. Using the free-field representation of the algebras $\widehat{\mathfrak{sl}}(2)_{1}$, $\widehat{\mathfrak{sl}}(2)_{2}$ and $\widehat{\mathfrak{sl}}(3)_{1}$ and restricting only on some components of $\mathcal{M}$ this table can be rewritten as
\begin{equation}\label{scheme-2}
 \begin{picture}(300,80)(0,40)
    \Thicklines
    \unitlength 1.6pt
    \put(0,0){\line(1,0){200}}
    \put(0,20){\line(1,0){200}}
    \put(0,40){\line(1,0){200}}
    \put(0,60){\line(1,0){200}}
    \put(0,0){\line(0,1){75}}
    \put(60,0){\line(0,1){75}}
    \put(120,0){\line(0,1){75}}
    \put(180,0){\line(0,1){75}}
    \put(27,8){\mbox{\small{$\mathcal{H}$}}}
    \put(77,8){\mbox{\small{$\mathcal{H}\oplus\mathsf{Vir}$}}}
    \put(137,8){\mbox{\small{$\mathcal{H}\oplus\mathsf{W}_{3}$}}}
    \put(20,28){\mbox{\small{$ \mathcal{H} \oplus \mathcal{H}$}}}
    \put(13,48){\mbox{\small{$ \mathcal{H} \oplus \mathcal{H}\oplus \mathcal{H}$}}}
    \put(62,28){\mbox{\small{$\mathcal{H}\oplus\mathcal{H}\oplus\mathcal{F}\oplus \textsf{NSR}$}}}
    \put(-22,8){\mbox{\small{$p=1$}}}
    \put(-22,28){\mbox{\small{$p=2$}}}
    \put(-22,48){\mbox{\small{$p=3$}}}
    \put(22,-8){\mbox{\small{$r=1$}}}
    \put(82,-8){\mbox{\small{$r=2$}}}
    \put(142,-8){\mbox{\small{$r=3$}}}
    \put(25,70){\circle*{1}}
    \put(30,70){\circle*{1}}
    \put(35,70){\circle*{1}}
    \put(85,50){\circle*{1}}
    \put(90,50){\circle*{1}}
    \put(95,50){\circle*{1}}
    \put(85,70){\circle*{1}}
    \put(90,70){\circle*{1}}
    \put(95,70){\circle*{1}}
    \put(145,30){\circle*{1}}
    \put(150,30){\circle*{1}}
    \put(155,30){\circle*{1}}
    \put(145,50){\circle*{1}}
    \put(150,50){\circle*{1}}
    \put(155,50){\circle*{1}}
    \put(145,70){\circle*{1}}
    \put(150,70){\circle*{1}}
    \put(155,70){\circle*{1}}
    \put(195,50){\circle*{1}}
    \put(200,50){\circle*{1}}
    \put(205,50){\circle*{1}}
    \put(195,30){\circle*{1}}
    \put(200,30){\circle*{1}}
    \put(205,30){\circle*{1}}
    \put(195,10){\circle*{1}}
    \put(200,10){\circle*{1}}
    \put(205,10){\circle*{1}}
    \end{picture}
\vspace*{1.9cm}
\end{equation}
\end{subequations}
where $\mathcal{F}$ is the Majorana fermion algebra.

In the language of the scheme \eqref{scheme} the conjecture of \cite{Belavin:2011pp}  imply that there exists a construction of geometrical action of the algebra \eqref{BF-coset} on equivariant cohomologies of $\mathcal{M}=\bigsqcup_{N}\mathcal{M}(r,N)^{\mathbb{Z}_{p}}$. This action was constructed explicitly only in the case of rank one ($r=1$) in \cite{0970.17017}. For higher ranks $r>1$ a similar construction is not developed so far. However, it can be obtained as a limit of geometrical action of more general algebra constructed by Nakajima in  \cite{Nakajima:fk}.  To be more precise, the author in  \cite{Nakajima:fk} constructed  the action of the so called  $\mathfrak{gl}_p$-toroidal algebra of the level $r$ on equivariant $K$-theory of the space  $\mathcal{M}=\bigsqcup_{N}\mathcal{M}(r,N)^{\mathbb{Z}_{p}}$. In some limit equivariant $K$-theory degenerates to equivariant cohomology and toroidal algebra degenerates to the Vertex operator algebra related to the coset $\mathcal{A}(r,p)$\footnote{Algebraic construction of such limit of toroidal algebra is given in the case $r=1$ \cite{10.1063/1.2823979,2010arXiv1002.2485F}, for $r>1$ \cite{Feigin-unpub}. The geometrical interpretation of the obtained coset algebras is very implicit.}. The construction based on a limit of toroidal algebra is difficult to accomplish (for $p=1$ case see \cite{Awata:2011fk}). However, using geometrical intuition one can predict the properties of the basis quoted above. It gives the expressions for the conformal blocks which can be compared to the expressions obtained from the standard CFT framework. Below we list main up-to-date achievements in this direction.
\begin{itemize}
\item In the case $p=1$, $r=1$ Nakajima \cite{Nakajima_1995} defined the geometrical action of the Heisenberg algebra. The fixed points basis corresponds to Jack polynomials, see e.g. \cite{Li:uq}. Carlsson and Okounkov gave geometrical construction of the vertex operator in \cite{Carlsson:2008fk}.
\item The case $p=1$, $r=2$ was considered in the paper \cite{Alday:2009aq}. The authors conjectured the expression for the multipoint conformal blocks in terms of the Nekrasov instanton partition functions. Alday and Tachikawa in \cite{Alday:2010vg} conjectured the existence of the basis which explains these expressions. In \cite{Alba:2010qc} explicit algebraic construction of this basis was given.
\item The case $p=1$, $r>2$ was considered along the lines of  \cite{Alday:2009aq} by Wyllard  \cite{Wyllard:2009hg} (see also \cite{Mironov:2009by}). The construction of the basis was done in \cite{Fateev:2011hq}.
\item For the case $p=2$, $r=2$ V.~Belavin and the third author proposed an expression for Whittaker limit of the four-point superconformal block in Neveu-Schwarz sector in terms of Nekrasov instanton partition functions \cite{Belavin:2011pp}. This result was generalized in \cite{Belavin:2011tb} for general four-point conformal block. For the results in Ramond sector see \cite{Ito:2011mw}.
\item For $p>2$.    The check of central charges of the coset CFT $\widehat{\mathfrak{sl}}(r)_p \times \widehat{\mathfrak{sl}}(r)_{n-p}\Bigl/\widehat{\mathfrak{sl}}(r)_n$ from $M-$theory consideration was performed in \cite{Nishioka:2011jk}.  Wyllard \cite{Wyllard:2011mn} considered the Whittaker limit  in the case $p=4$, $r=2$.   Some further checks for this case  were made in \cite{Alfimov:2011ju}. In the case of generic  $p$ and $r$ some non-trivial checks were done in \cite{Wyllard:2011mn} by use of Kac determinant of the coset CFT.
\end{itemize}

There exists another compactification of the space of instantons on $\mathbb{C}^2/\mathbb{Z}_p$. Denote by $X_p$ the minimal resolution of the $\mathbb{C}^2/\mathbb{Z}_p$. The moduli space $\mathcal{M}(X_{2},r,N)$ of framed torsion free sheaves of rank $r$  on $X_p$ is a smooth compactification of the space of instantons on $\mathbb{C}^2/\mathbb{Z}_p$. The torus action on $\mathcal{M}(X_{2},r,N)$ is  induced by the  torus action on $X_p$ and action on framing at infinity. The fixed points are labelled by $p$ sets  of $r$-tuple of Young diagrams and $p-1$  vectors $(k^{i}_1,k^{i}_2\dots,k^{i}_r)$, $1 \leq i \leq p-1$ of integer numbers. Note that this combinatorial description differs from the description for torus fixed points on $\mathcal{M}(r,N)^{\mathbb{Z}_p}$ in terms of $p$-colors colored Young diagrams. It is natural to assume that similar algebras act on the equivariant cohomologies of $\mathcal{M}(X_{2},r,N)$. In \cite{Bonelli:2011jx,Bonelli:2011kv} the authors used the space $\mathcal{M}(X_{2},2,N)$ for Nekrasov type expressions of the conformal blocks in the superconformal field theory.

The symmetry algebra for the  coset models
\begin{equation}\label{BF-coset-2}
  \frac{\widehat{\mathfrak{sl}}(r)_p \times \widehat{\mathfrak{sl}}(r)_{n-p}}{\widehat{\mathfrak{sl}}(r)_n}
\end{equation}
with generic $r$ and $p$ is not known in explicit form. For example for $r=2$ and generic $p$  the symmetry algebra is generated by the current $G(z)$ of fractional spin $(p+4)/(p+2)$ \cite{Argyres:1990aq}. This current is non-abelianly braided i.e. the operator product of $G(z)$ with itself contains singularities with incommensurable powers. This fact makes it difficult to study such models. The situation simplifies in three cases: $p=1$ which corresponds to the Virasoro algebra, $p=2$ which corresponds to the Neveu-Schwarz-Ramond algebra and $p=4$ which  can be expressed through the abelianly braided model called spin $4/3$ parafermionic CFT  \cite{Fateev:1985ig,Pogosian:1988ar}. For higher ranks the algebraic treatment of the coset model \eqref{BF-coset-2} becomes even more problematic. Already in the case of $p=1$ the commutation relations of the corresponding algebra ($\mathsf{W}_{r}$ algebra in this case) are known in explicit terms only for the small ranks. Remarkably, that such obstructions do not appear in geometrical side of the relation \eqref{scheme} and the case of generic $p$ and $r$ can be studied in its entirety.

In this paper we continue the study of the case $p=2$, $r=2$ as the next example (after $p=1$ and $r=2$) where the algebraic treatment is relatively simple\footnote{Some analysis of the case $p=4$ and $r=2$  was done in \cite{Wyllard:2011mn,Alfimov:2011ju}.}. General philosophy suggests the existence of the  basis in the representation of the algebra $\mathcal{H}\oplus\mathcal{H}\oplus\mathcal{F}\oplus \textsf{NSR}$ (see \eqref{scheme-2}). This basis has geometric origin and gives expressions for the conformal blocks mentioned before. Moreover, the different manifolds $\mathcal{M}(X_{2},2,N)$ and $\mathcal{M}(2,N)^{\mathbb{Z}_2}$ might correspond to different bases.

The appearance of the different bases is a new effect in the case $p>1$ compared to $p=1$. Geometrically this is related to the fact that manifolds $\mathcal{M}(X_{2},2,N)$ and $\mathcal{M}(2,N)^{\mathbb{Z}_2}$ are $\mathbb{C}^*$-- diffeomorphic, but not $\left(\mathbb{C}^*\right)^2$-- diffeomorphic. Algebraically this leads to the fact that formulae in \cite{Belavin:2011pp,Belavin:2011tb} from the one hand and \cite{Bonelli:2011jx,Bonelli:2011kv} from the other hand are different. They give the same result because the manifolds $\mathcal{M}(X_{2},2,N)$ and $\mathcal{M}(2,N)^{\mathbb{Z}_2}$ are the compactifications of the same manifold and hence the integrals are equal. In other words these two compactifications give two ways to compute the integral. Equality between results means the nontrivial combinatorial identity.

In section \ref{SAGT} we construct the basis which corresponds to the manifold $\mathcal{M}_2(2,N)$ (to be more precise to its component with $c_1=0$). This basis gives \cite{Bonelli:2011jx,Bonelli:2011kv} expressions for the  conformal blocks in the superconformal field theory. As the main tool we use the subalgebra
$$
\left(\mathcal{H} \oplus \textsf{Vir} \right) \oplus \left(\mathcal{H} \oplus \textsf{Vir} \right) \subset
\left(\mathcal{H}\oplus\mathcal{H}\oplus\mathcal{F}\oplus\mathsf{NSR}\right).
$$
In other words we use an embedding of the  direct sum of two algebras for $p=1$ into the algebra for $p=2$ (see \eqref{scheme-2}). Geometrically the appearance of this subalgebra is related to the existence of two  points on $X_{2}$  invariant under the torus action. Algebraic explanation based on the coset formula
$$\frac{\widehat{\mathfrak{gl}}(n)_r}{\widehat{\mathfrak{gl}}(n-1)_r}
\times \frac{\widehat{\mathfrak{gl}}(n-1)_r}{\widehat{\mathfrak{gl}}(n-2)_r} \subset \frac{\widehat{\mathfrak{gl}}(n)_r}{\widehat{\mathfrak{gl}}(n-2)_r}.
$$
Using this subalgebras we reduce the basis problem to the $p=1$ case and use  construction of \cite{Alba:2010qc}.

In section \ref{SAGT-collored} we study the basis corresponding to the manifold $\mathcal{M}(2,N)^{\mathbb{Z}_2}$ (to be more precise only one connected component for each $N$). We couldn't give an explicit construction of this basis but we conjecture a factorized formula for matrix elements of vertex operators ($Z_{\textsf{bif}}$) in this basis. We checked this formula comparing two evaluations of the five-point conformal block. In the first case we use  the formula mentioned above connected with the hypothetical basis which corresponds  to the manifold $\mathcal{M}(2,N)^{\mathbb{Z}_{2}}$.
In the second case we use the basis constructed in section \ref{SAGT}. This basis corresponds to the manifold $\mathcal{M}(X_{2},2,N)$.

In the second part of section \ref{SAGT-collored} we study all connected components of $\mathcal{M}(1,N)^{\mathbb{Z}_2}$. In other words it means that we consider the algebra $\mathcal{H}\oplus\widehat{\mathfrak{sl}}(2)_{1}$ from the table \eqref{scheme-1} instead of the algebra $\mathcal{H}\oplus\mathcal{H}$ from the table \eqref{scheme-2}. We will see that there are several classes of connected components labeled by an integer number $d$ and different classes correspond to different bases. The basis constructed in section \ref{SAGT} appears to be a limit when $d \rightarrow \infty$.

The plan of the paper is the following. In section \ref{AFLT} we reproduce all known facts about the basis in the case $p=1$. The content of the  sections \ref{SAGT} and \ref{SAGT-collored} was described above.   In \ref{Concl} we formulate some obvious open questions. In appendix \ref{2Liouville} we discuss the embedding $\mathsf{Vir}\oplus\mathsf{Vir}\subset\mathcal{F}\oplus\mathsf{NSR}$ in more details. In appendices \ref{Highest-weight} and \ref{Bersh-app} we present some explicit formulae used in sections \ref{SAGT} and \ref{SAGT-collored}.
\section{The case $p=1$}\label{AFLT}
In this section we review the construction of the basis in the case $p=1$ and arbitrary rank $r$. This example is used to illustrate the general scheme formulated in Introduction. Moreover, some constructions  will be used below in section \ref{SAGT}.
\subsection{Geometrical setup}
In this case the geometrical object under consideration is the manifold $\mathcal{M}=\bigsqcup_{N}\mathcal{M}(r,N)$, where $\mathcal{M}(r,N)$ is the compactified moduli spaces of $U(r)$ instantons on $\mathbb{C}^{2}$  with instanton number $N$ (see \cite{0949.14001} Ch. 2 or \cite{2003math.....11058N} Ch. 3)
\begin{equation}\label{ADHM-def}
  \mathcal{M}(r,N)\cong\left\{
  (B_{1},B_{2},I,J)\left|
  \begin{aligned}
  &(\mathrm{i})\quad[B_{1},B_{2}]+IJ=0\\
  &(\mathrm{ii})\quad
  \begin{minipage}{.44\textwidth}
 There is no subspace $S \varsubsetneq \mathbb{C}^n$, such that $B_\sigma S \subset S$ ($\sigma= 1,2$) and $I_1,\dots I_r \in S$
  \end{minipage}¥
  \end{aligned}
  \right\}\right.\Biggl/\mathrm{GL_{N}},
\end{equation}
where $B_{j}$,  $I$ and $J$ are $N\times N$, $N\times r$ and $r\times N$ complex matrices with the action of $\mathrm{GL}_{N}$ given by
\begin{equation*}
   g\cdot(B_{1},B_{2},I,J)=(gB_{1}g^{-1},gB_{2}g^{-1},gI,Jg^{-1}),
\end{equation*}
for $g\in\mathrm{GL}_{N}$. In \eqref{ADHM-def} $I_{1},\dots,I_{r}$ denote the columns of the matrix $I$.
Torus $T=(\mathbb{C}^*)^2\times (\mathbb{C}^*)^{r}$ acts on the manifold $\mathcal{M}$. The action of  $(\mathbb{C}^*)^2$ arise from the action of two  rotations on $\mathbb{C}^2$ and   $(\mathbb{C}^*)^r$ action arises from the action on framing at infinity. The exact formula reads
\begin{equation}\label{torus-action}
B_1\mapsto t_1 B_1 ; \, \, \, \, B_1\mapsto t_1 B_1 ; \, \, \, \, I \mapsto It; \, \, \,
\, J\mapsto t_1 t_2 t^{-1}J,
\end{equation}
where $(t_1,t_2,t)\in \mathbb{C}^*\times \mathbb{C}^*\times (\mathbb{C}^*)^r=T$. Fixed points under the torus action are labeled by the  $r$-tuples of Young diagrams $\vec{Y}=(Y_1,\dots,Y_{r})$ and $T$ acts on the tangent space of any fixed point $p_{\scriptscriptstyle{\vec{Y}}}=p_{\scriptscriptstyle{Y_1},\dots,\scriptscriptstyle{Y_r}}$. For any element $v=(\epsilon_1,\epsilon_2,a)\in\textit{Lie}(T)$, where $\epsilon_{1},\epsilon_{2}\in\mathbb{C}$,  $a$ is the diagonal matrix $a=\textrm{diag}(a_{1},\dots,a_{r})$ and the determinant of $v$ on the tangent space of $p_{\scriptscriptstyle{\vec{Y}}}$ reads \cite{Flume:2002az,2003math......6198N}
\begin{equation}\label{det-vp}
 \det v\Bigl|_{p_{\scriptscriptstyle{\vec{Y}}}}=\prod_{i,j=1}^{r}
 \prod_{s\in \scriptscriptstyle{Y_{i}}}
    E_{\scriptscriptstyle{Y_{i}},\scriptscriptstyle{Y_{j}}}(a_{i}-a_{j}|s)
   \bigl(\epsilon_{1}+\epsilon_{2}-E_{\scriptscriptstyle{Y_{i}},\scriptscriptstyle{Y_{j}}}(a_{i}-a_{j}|s)\bigr),
\end{equation}
where
\addtocounter{equation}{-1}
\begin{subequations}
\begin{equation}\label{E-def}
    E_{\scriptscriptstyle{Y},\scriptscriptstyle{W}}(x|s)=x-\epsilon_{1}\,\mathrm{l}_{\scriptscriptstyle{W}}(s)+\epsilon_{2}(\mathrm{a}_{\scriptscriptstyle{Y}}(s)+1).
\end{equation}
\end{subequations}
In \eqref{E-def} $\mathrm{a}_{\scriptscriptstyle{Y}}(s)$ and $\mathrm{l}_{\scriptscriptstyle{W}}(s)$ are correspondingly the arm length of the box $s$ in the partition $Y$ and the leg length of the box $s$ in the partition $W$. The inverse of  the determinant \eqref{det-vp} usually called the contribution of the vector hypermultiplet and denoted as
\begin{equation}\label{Zvec-def}
    Z^{(r)}_{\textsf{vec}}(\vec{a},\vec{Y}|\epsilon_{1},\epsilon_{2})\overset{\text{def}}{=}
    \prod_{i,j=1}^{r}
 \prod_{s\in \scriptscriptstyle{Y_{i}}}
    \Bigl(E_{\scriptscriptstyle{Y_{i}},\scriptscriptstyle{Y_{j}}}(a_{i}-a_{j}|s)
   \bigl(\epsilon_{1}+\epsilon_{2}-E_{\scriptscriptstyle{Y_{i}},\scriptscriptstyle{Y_{j}}}(a_{i}-a_{j}|s)\bigr)\Bigr)^{-1},
\end{equation}
where $\vec{a}=(a_{1},\dots,a_{r})$.
This quantity enters into instanton part of the Nekrasov partition function for pure $U(r)$ gauge theory (without matter)
\begin{equation}\label{Zpure-def}
   Z^{(r)}_{\text{pure}}(\vec{a},\epsilon_{1},\epsilon_{2}|\Lambda)=
   1+\sum_{k=1}^{\infty}\sum_{|\scriptscriptstyle{\vec{Y}}\scriptstyle|=k} Z^{(r)}_{\textsf{vec}}(\vec{a},\vec{Y}|\epsilon_{1},\epsilon_{2})\,\Lambda^{4k},
\end{equation}
where $\vec{a}=(a_{1},\dots,a_{r})$ is interpreted as vacuum expectation value of the scalar field and $\Lambda$ is the scale in gauge theory.

An important quantity is the contribution of the bifundamental matter hypermultiplet \cite{Fucito:2004gi,Flume:2002az,Shadchin:2005cc}. This quantity is defined geometrically and is given by the determinant of the vector field in a fiber of certain bundle over fixed point\footnote{This fixed point is labeled by the pair of $r-$tuples of Young diagrams $\vec{Y}$ and $\vec{W}$.} of the torus on $\mathcal{M}(r,N)\times\mathcal{M}(r,N')$
\begin{equation}\label{Zbif-def}
    Z^{(r)}_{\text{\sf{bif}}}(m;\vec{a}',\vec{W};\vec{a},\vec{Y}|\epsilon_{1},\epsilon_{2})=\prod_{i,j=1}^{r}
    \prod_{s\in \scriptscriptstyle{Y_{i}}}\left(\epsilon_{1}+\epsilon_{2}-E_{\scriptscriptstyle{Y_{i}},\scriptscriptstyle{W_{j}}}(a_{i}-a'_{j}|s)-m\right)
    \prod_{t\in \scriptscriptstyle{W_{j}}}\left(E_{\scriptscriptstyle{W_{j}},\scriptscriptstyle{Y_{i}}}(a'_{j}-a_{i}|t)-m\right),
\end{equation}
where the  parameter $m$ coincides with the mass of bifundamental hypermultiplet.
As all the expressions $Z_{\textsf{vec}}$ and $Z_{\textsf{bif}}$ appear to be homogeneous under $a_{i}\rightarrow\lambda a_{i}$,  $m\rightarrow\lambda m$ and $\epsilon_{j}\rightarrow\lambda\epsilon_{j}$ one can fix this freedom by demanding that $\epsilon_{1}\epsilon_{2}=1$. We will adopt the notations common in CFT literature
\begin{equation*}
   \epsilon_{1}=b,\qquad
   \epsilon_{2}=b^{-1}.
\end{equation*}
Moreover, we assume that $\sum_{j=1}^{r}a_{j}=0$. In particular, below we consider in details the case $r=1$ and $r=2$. For $r=2$ it would be convenient to introduce
\begin{equation}\label{Zbif-def-2}
  \mathbb{F}(\alpha|P',\vec{W};P,\vec{Y})\overset{\text{def}}{=}
  Z^{(2)}_{\text{\sf{bif}}}(\alpha;(P',-P'),\vec{W};(P,-P),\vec{Y}|b,1/b).
\end{equation}
and
\begin{equation}\label{Nvec-def-2}
   \mathbb{N}(P,\vec{Y})\overset{\text{def}}{=}
   Z_{\textsf{vec}}^{(2)}((P,-P),\vec{Y}|b,1/b).
\end{equation}
\subsection{Algebraic setup}
In this case the conformal field theory under consideration has the symmetry algebra $\mathcal{H}\oplus\textsf{W}_{r}$. There is special basis of states in the highest weight representation of this algebra corresponding to the fixed points of the vector field acting on $\mathcal{M}$. This basis of states diagonalizes an infinite system of commuting quantities (Integrals of Motion) $\mathbf{I}_{k}$
\begin{equation}\label{II-commute}
   [\mathbf{I}_{k},\mathbf{I}_{l}]=0,
\end{equation}
which are elements of the universal enveloping of the algebra $\mathcal{H}\oplus\textsf{W}_{r}$.
We review the construction of the basis of states in two particular cases $r=1$ and $r=2$. For the case of general rank see \cite{Fateev:2011hq}.
\subsubsection{Case $r=1$}
Our algebra is Heisenberg algebra with components $\mathtt{a}_{k}$ and commutation relations\footnote{Here and below we assume that our Heisenberg  algebra  has no zero mode since it plays artificial role in our construction. In other words we assume that we are considering highest weight representations such that $a_{0}|0\rangle=0$.}
\begin{equation}\label{H-relat}
  [\mathtt{a}_{n},\mathtt{a}_{m}]=n\,\delta_{n+m,0}.
\end{equation}
The highest weight representation of this algebra  (Fock module) is defined by the vacuum state $|0\rangle$
\begin{equation*}
   \mathtt{a}_{n}|0\rangle=0\quad\text{for}\quad n>0,
\end{equation*}
and spanned by the vectors of the form
\begin{equation}
   \mathtt{a}_{-k_{1}}\dots\mathtt{a}_{-k_{n}}|0\rangle,\qquad
   k_{1}\geq k_{2}\geq\dots\geq k_{n}.
\end{equation}
One can define another basis
\begin{equation}\label{Jack-basis-1}
   |Y\rangle\overset{\text{def}}{=}\jac_{\scriptscriptstyle{Y}}^{\scriptscriptstyle{(1/g)}}(x)|0\rangle,
\end{equation}
where $\jac_{\scriptscriptstyle{Y}}^{\scriptscriptstyle{(1/g)}}(x)$ is the Jack polynomial in integral normalization \cite{Macdonald} with parameter $g=-b^{2}$ associated to the partition $Y$ and the following identification is made
\begin{equation*}
    \mathtt{a}_{-k}=-ib\,p_{k},
\end{equation*}
where $p_{k}$ are power-sum symmetric polynomials
\begin{equation*}
   p_{k}=p_{k}(x)=\sum_{j}x_{j}^{k}.
\end{equation*}
The basis of states $|Y\rangle$ is usually called Jack basis by transparent reasons. There exists a system of Integrals of Motion $\mathbf{I}_{k}$  which acts diagonally in Jack basis \eqref{Jack-basis-1}. The first two representatives of this family are (here $Q=b+1/b$)
\begin{equation}\label{I21-components}
 \begin{aligned}
   &\mathbf{I}_{1}=\sum_{k>0}\mathtt{a}_{-k}\mathtt{a}_{k},\\
   &\mathbf{I}_{2}=iQ\sum_{k>0}k\mathtt{a}_{-k}\mathtt{a}_{k}+\frac{1}{3}\sum_{i+j+k=0}\mathtt{a}_{i}\mathtt{a}_{j}\mathtt{a}_{k}.
 \end{aligned}
\end{equation}

Another important property of the Jack basis was pointed out in \cite{Carlsson:2008fk}. Namely, consider vertex operator
\begin{equation}\label{vertex-CO}
   \mathsf{V}_{\alpha}=
   e^{(\alpha-Q)\varphi_{-}(1)}e^{\alpha\varphi_{+}(1)},
\end{equation}
with $\varphi_{+}(z)=i\sum_{n>0}\frac{\mathtt{a}_{n}}{n}z^{-n}$ and $\varphi_{-}(z)=i\sum_{n<0}\frac{\mathtt{a}_{n}}{n}z^{-n}$. Define also dual basis $\langle W|$, which is orthogonal to the Jack basis with respect to usual scalar product in the Heisenberg algebra. It was proved in \cite{Carlsson:2008fk} that
\begin{equation}\label{matrix-element-CO}
  \langle W|\mathsf{V}_{\alpha}|Y\rangle=
  \prod_{s\in \scriptscriptstyle{Y}}
  \Bigl(b\,\bigl(\mathrm{l}_{\scriptscriptstyle{W}}(s)+1\bigr)-b^{-1}\mathrm{a}_{\scriptscriptstyle{Y}}(s)-\alpha\Bigr)
  \prod_{t\in \scriptscriptstyle{W}}
  \Bigl(b^{-1}\,\bigl(\mathrm{a}_{\scriptscriptstyle{W}}(t)+1\bigr)-b\,\mathrm{l}_{\scriptscriptstyle{Y}}(t)-\alpha\Bigr).
\end{equation}

We stress that the Jack basis $|Y\rangle$ is interpreted as the basis of fixed points $p_{\scriptscriptstyle{Y}}$ of the vector field on instanton manifold $\mathcal{M}$ (in the case of rank one and $\epsilon_{1}=b$, $\epsilon_{2}=1/b$)\cite{Li:uq}. Integrals of Motion are interpreted as operators of multiplication on cohomology classes. We note that the r.h.s. of \eqref{matrix-element-CO} coincides with \eqref{Zbif-def} in the case of $r=1$, $a=a'=0$, $m=\alpha$ and $\epsilon_{1}=b$, $\epsilon_{2}=1/b$.
\begin{equation*}
  \langle W|\mathsf{V}_{\alpha}|Y\rangle=
   Z^{(1)}_{\text{\sf{bif}}}(\alpha;0,W;0,Y|b,b^{-1})
\end{equation*}

\subsubsection{Case $r=2$}
We consider conformal field theory, whose symmetry algebra is  $\mathcal{A}=\mathcal{H}\oplus\text{\sf Vir}$ (we use conventions which are specific in this case: there is the factor $1/2$ in commutation relations for $a_{k}$ generators compared to \eqref{H-relat})
\begin{equation}\label{Vir-relat}
   \begin{aligned}
   &[L_{n},L_{m}]=(n-m)L_{n+m}+\frac{c}{12}(n^{3}-n)\,\delta_{n+m,0},\\
   &[a_{n},a_{m}]=\frac{n}{2}\,\delta_{n+m,0},\qquad [L_{n},a_{m}]=0.
   \end{aligned}
\end{equation}
We will parametrize the central charge $c$ of the Virasoro algebra in a Liouville manner as
\begin{equation}
   c=1+6Q^{2},\qquad\text{where}\quad Q=b+\frac{1}{b}.
\end{equation}
We also need to introduce the  operators
\begin{equation}\label{primary}
    V_{\alpha}\overset{\text{def}}{=}\mathcal{V}_{\alpha}\cdot V_{\alpha}^{\scriptscriptstyle{\textsf{Vir}}},
\end{equation}
where $V_{\alpha}^{\scriptscriptstyle{\textsf{Vir}}}$ is the primary field of the Virasoro algebra with conformal dimension
\begin{equation}\label{Delta-Vir}
  \Delta(\alpha,b)=\alpha(Q-\alpha)
\end{equation}
and $\mathcal{V}_{\alpha}$ is a free exponential
\begin{equation}\label{vertex}
    \mathcal{V}_{\alpha}=e^{2(\alpha-Q)\varphi_{-}}e^{2\alpha\varphi_{+}},
\end{equation}
with $\varphi_{+}(z)=i\sum_{n>0}\frac{a_{n}}{n}z^{-n}$ and $\varphi_{-}(z)=i\sum_{n<0}\frac{a_{n}}{n}z^{-n}$.

Let us consider the highest weight representation of the algebra $\mathcal{H}\oplus\mathsf{Vir}$ parameterized by the momenta $P$ and defined by the vacuum state $|P\rangle$:
\begin{equation*}
    L_{n}|P\rangle=a_{n}|P\rangle=0,\quad\text{for}\quad n>0,\qquad
    L_{0}|P\rangle=\Delta(P)|P\rangle,\qquad \langle P|P\rangle=1.
\end{equation*}
The Virasoro conformal dimension of the state $|P\rangle$ is expressed through the momenta $P$ as
\begin{equation*}
\Delta(P)=\frac{Q^{2}}{4}-P^{2}.
\end{equation*}
Then the highest weight representation is spanned by the vectors of the form
\begin{equation}\label{naive-basis}
\begin{gathered}
   a_{-l_{m}}\dots a_{-l_{1}}L_{-k_{n}}\dots L_{-k_{1}}|P\rangle,\\
    k=(k_{1}\geq k_{2}\geq\dots\geq k_{n}),\quad
    l=(l_{1}\geq l_{2}\geq\dots\geq l_{m}).
\end{gathered}
\end{equation}
This representation is irreducible for general values of the momenta $P$.

In principle, one can choose another basis different from the naive one \eqref{naive-basis}. Among the possible bases there is one which is of special interest for us. The defining property of this basis is formulated by the following proposition proved in \cite{Alba:2010qc}.
\begin{Proposition}\label{AFLT-prop}
There exists unique orthogonal basis $|P\rangle_{\vec{\scriptscriptstyle{Y}}}$ such that
\begin{equation}\label{matrix-elements}
    \frac{ _{\vec{\scriptscriptstyle{W}}}\langle P'|V_{\alpha}|P\rangle_{\vec{\scriptscriptstyle{Y}}}}
    {\langle P'|V_{\alpha}|P\rangle}=\mathbb{F}(\alpha|P',\vec{W};P,\vec{Y}).
\end{equation}
\end{Proposition}
In proposition \ref{AFLT-prop} we denoted the elements of this basis by $|P\rangle_{\scriptscriptstyle{\vec{Y}}}$ where $\vec{Y}=(Y_{1},Y_{2})$ stands for the pair of Young diagrams. In \eqref{matrix-elements} the function $\mathbb{F}(\alpha|P',\vec{Y}';P,\vec{Y})$ is defined by \eqref{Zbif-def}--\eqref{Zbif-def-2}. We note that in geometrical language the basis state $|P\rangle_{\vec{\scriptscriptstyle{Y}}}$ corresponds to the fixed point $p_{\vec{\scriptscriptstyle{Y}}}$ of the vector field. It follows from Proposition \ref{AFLT-prop} that the states $|P\rangle_{\vec{\scriptscriptstyle{Y}}}$ form an orthogonal basis
\begin{equation}
    _{\vec{\scriptscriptstyle{W}}}\langle P|P\rangle_{\vec{\scriptscriptstyle{Y}}}=
    \frac{\delta_{\vec{\scriptscriptstyle{Y}},\vec{\scriptscriptstyle{W}}}}{\mathbb{N}(P,\vec{Y})},
\end{equation}
where $\delta_{\vec{\scriptscriptstyle{Y}},\vec{\scriptscriptstyle{W}}}=0$ if $\vec{Y}\neq\vec{W}$, $\delta_{\vec{\scriptscriptstyle{Y}},\vec{\scriptscriptstyle{Y}}}=1$ and function $\mathbb{N}(P,\vec{Y})$ is defined by \eqref{Nvec-def-2}.

It will be convenient below to introduce  operators $X_{\vec{\scriptscriptstyle{Y}}}(P,b)$:
\begin{equation}\label{X-def}
  |P\rangle_{\vec{\scriptscriptstyle{Y}}}\overset{\text{def}}{=}X_{\vec{\scriptscriptstyle{Y}}}(P,b)|P\rangle,
\end{equation}
and such that $X_{\vec{\scriptscriptstyle{Y}}}(P,b)$ does not contain positive components of $\mathcal{A}$, i.e.
\begin{equation}
  X_{\vec{\scriptscriptstyle{Y}}}(P,b)=\sum_{\scriptscriptstyle{l+k}=|\scriptscriptstyle{Y}|}C_{\vec{\scriptscriptstyle{Y}}}^{\scriptscriptstyle{\vec{l}},\scriptscriptstyle{\vec{k}}}(P,b)\,a_{-l_{m}}\dots a_{-l_{1}}L_{-k_{n}}\dots L_{-k_{1}},
\end{equation}
where $l=\sum l_{i}$ and $k=\sum k_{j}$.
It can be shown that all the  coefficients $C_{\vec{\scriptscriptstyle{Y}}}^{\scriptscriptstyle{\vec{l}},\scriptscriptstyle{\vec{k}}}(P,b)$ are some polynomials in the momenta $P$ (see examples in \cite{Alba:2010qc}).

The system of Integrals of Motion which acts diagonally in the basis  $|P\rangle_{\vec{\scriptscriptstyle{Y}}}$ was constructed in \cite{Alba:2010qc}. First two representatives of this system are
\begin{equation}\label{I1I2}
  \begin{aligned}
     &\mathbf{I}_{1}=L_{0}+2\sum_{k>0}a_{-k}a_{k},\\
     &\mathbf{I}_{2}=
     \sum_{k\neq0}a_{-k}L_{k}+2iQ\sum_{k>0}^{\infty}ka_{-k}a_{k}+\frac{1}{3}\sum_{i+j+k=0}a_{i}a_{j}a_{k}.
  \end{aligned}
\end{equation}
This integrable system was studied in \cite{Alba:2010qc,Belavin:2011js,Estienne:2011qk}. In particular, it was noticed that the basis of eigenstates is very similar to the Jack basis studied above. The states $|P\rangle_{\scriptscriptstyle{Y,\varnothing}}$ as well as the states $|P\rangle_{\scriptscriptstyle{\varnothing,Y}}$ become the Jack states \eqref{Jack-basis-1}  if one expresses  the Virasoro generators $L_{n}$ in terms of bosons. In fact, there are two ways to do it
\begin{equation}\label{Bosonization}
  \begin{gathered}
   L_{n}=\sum_{k\neq0,n}c_{k}c_{n-k}+i(nQ\mp2\mathcal{P})c_{n},\quad L_{0}=\frac{Q^{2}}{4}-\mathcal{P}^{2}+2\sum_{k>0}c_{-k}c_{k},\\
    [c_{n},c_{m}]=\frac{n}{2}\,\delta_{n+m,0},\quad[\mathcal{P},c_{n}]=0,\quad\mathcal{P}|P\rangle=P|P\rangle,\quad\langle P|\mathcal{P}=-P\langle P|.
  \end{gathered}
\end{equation}
corresponding to the choice of sign in front of operator of the zero mode $\mathcal{P}$.  These two choices define two different sets of bosons $c_{k}$, which are related by the unitary transform also called reflection operator \cite{Zamolodchikov:1995aa}. The sign ``$-$'' works for the states  $|P\rangle_{\scriptscriptstyle{Y,\varnothing}}$ while ``$+$'' works for $|P\rangle_{\scriptscriptstyle{\varnothing,Y}}$. For example, taking ``$-$'' in \eqref{Bosonization} one can show that
\begin{equation}\label{Jack-basis}
    |P\rangle_{\scriptscriptstyle{Y,\varnothing}}=\Omega_{\scriptscriptstyle{Y}}(P)\,\jac_{\scriptscriptstyle{Y}}^{\scriptscriptstyle{(1/g)}}(x)|P\rangle,
\end{equation}
where $\jac_{\scriptscriptstyle{Y}}^{\scriptscriptstyle{(1/g)}}(x)$ is the Jack polynomial with  $g=-b^{2}$,
\begin{equation*}
     a_{-k}-c_{-k}=-ib\,p_{k}(x),
\end{equation*}
and $\Omega_{\scriptscriptstyle{Y}}(P)$ is the normalization factor, whose explicit form can be found in \cite{Alba:2010qc}. The statement similar to \eqref{Jack-basis} is valid for the state $|P\rangle_{\scriptscriptstyle{\varnothing,Y}}$ if one takes the sign ``$+$'' in \eqref{Bosonization}. At the value $Q=0$ these two sets of bosons are differ by sign and general state $|P\rangle_{\scriptscriptstyle{\vec{Y}}}$ can be written as a tensor product of two Jack states \cite{Belavin:2011js}.
Remarkably, the fact that some of the states become the Jack states after bosonization is valid for any $r$ (see \cite{Fateev:2011hq}). Using this fact and the ``bootstrap'' equations suggested in \cite{Alba:2010qc,Fateev:2011hq} one can construct recurrently all basis states.
\section{Supersymmetric case ($p=2$, $r=2$)}\label{SAGT}
In this section we construct the basis corresponding to the case $p=2$, $r=2$ from the general scheme. In algebraic side we expect to deal with the algebra $\mathcal{A}=\mathcal{H}\oplus\mathcal{H}\oplus\mathcal{F}\oplus \textsf{NSR}$.
\subsection{Geometrical setup}

By $X_2$ we denote the ALE space, which is the minimal resolution of the factor space $\mathbb{C}^{2}/\mathbb{Z}_{2}$. This space can be constructed by gluing two charts $\mathbb{C}^2$ with coordinates:
$$
 \text{1}: \quad \mathbb{C}^2\; (u_1,v_1)\quad u_2=v_1^{-1},\,\; v_2=u_1v_1^2 \qquad\qquad
 \text{2}: \quad \mathbb{C}^2\; (u_2,v_2)\quad u_1=u_2^2v_2,\,\; v_1=u_2^{-1}
$$
There is a map $\mathbb{C}^2\backslash \{0\} \rightarrow X_2$ given in coordinates $u_1=z_1^2, v_1=z_2/z_1$ in the first chart and $u_2=z_1/z_2, v_2=z_2^2$ in the second chart. Points $(z_1,z_2)$ and $(-z_1,-z_2)$ have the same image under this map. Hence we obtain the projection $$\pi\colon X_2 \rightarrow \mathbb{C}^{2}/\mathbb{Z}_{2},$$
which appears to be the minimal resolution of singularity. The preimage of $(0,0)\in \mathbb{C}^2$ is exceptional divisor $C \in X_2$. In the first and the second charts $C$ is given by equations $u_1=0$ and $v_2=0$ respectively.

The torus action on $X_2$ arises from the torus action on $\mathbb{C}^2$:
$$ \text{1:}\quad (u_1,v_1)\mapsto (t_1^2u_1,t^{-1}_1t_2v_1); \qquad \text{2:}\quad (u_2,v_2)\mapsto (t_1t^{-1}_2u_2,t_2^2v_2).$$
There are two points which are invariant under the torus action namely $p_1$ and $p_2$ origins in the first and second charts respectively.

Let $\mathcal{M}=\bigsqcup_{N}\mathcal{M}(X_{2},2,N)$ be the moduli space of framed torsion free sheaves on $X_{2}$ of rank $2$ with Chern classes $c_{1}=0$, $c_{2}=N$ \cite{1166.14007}. Torus $T=(\mathbb{C}^*)^2\times (\mathbb{C}^*)^{2}$ acts on the manifold $\mathcal{M}$. The action of the first $(\mathbb{C}^*)^2$ arise from the action of two rotations on $\mathbb{C}^2$ and the action of the second $(\mathbb{C}^*)^r$ action arises from the action on framing at infinity.

The points of the torus were described in \cite{2011CMaPh.304..395B}. They are labeled by the pair of pairs of Young diagrams $\vec{Y}^{\scriptscriptstyle{(\sigma)}}=(Y_{1}^{\scriptscriptstyle{(\sigma)}},Y_{2}^{\scriptscriptstyle{(\sigma)}})$, $\sigma=1,2$ and one integer number $k\in\mathbb{Z}$. The pair of Young diagrams $\vec{Y}^{\scriptscriptstyle{(\sigma)}}$ describes the corresponding sheaf $\mathcal{E}_{\scriptscriptstyle\vec{Y}^{\scriptscriptstyle{(\sigma)}},k}$ near the invariant point $p_\sigma$ and $k$ means that $\mathcal{E}_{\scriptscriptstyle\vec{Y}^{\scriptscriptstyle{(\sigma)}},k}$ is a subsheaf of $\mathcal{O}(kC)+\mathcal{O}(-kC)$.

The determinant of the vector field $v=(\epsilon_{1},\epsilon_{2},a)$ at the fixed point $p_{\scriptscriptstyle\vec{Y}^{\scriptscriptstyle{(\sigma)}},k}$ equals to \cite{2011CMaPh.304..395B}
\begin{equation}\label{Zvec-def-2}
   \det v\Bigl|_{p_{\scriptscriptstyle\vec{Y}^{\scriptscriptstyle{(\sigma)}},k}}=\frac{l_{\vec{k}}(\vec{a}|\epsilon_{1},\epsilon_{2})}
   {Z_{\textsf{vec}}^{(2)}(\vec{a}+\epsilon_{1}\vec{k},\vec{Y}^{\scriptscriptstyle{(1)}}|2\epsilon_{1},\epsilon_{2}-\epsilon_{1})
   Z_{\textsf{vec}}^{(2)}(\vec{a}+\epsilon_{2}\vec{k},\vec{Y}^{\scriptscriptstyle{(2)}}|\epsilon_{1}-\epsilon_{2},2\epsilon_{2})},
\end{equation}
where $\vec{k}=(k,-k)$, function $Z_{\textsf{vec}}^{(2)}(\vec{a},\vec{Y}|\epsilon_{1},\epsilon_{2})$ is given by  \eqref{Zvec-def} and the factor $l_{\vec{k}}(\vec{a}|\epsilon_{1},\epsilon_{2})$ is
\begin{equation}\label{l-vec}
   l_{\vec{k}}(\vec{a}|\epsilon_{1},\epsilon_{2})=(-1)^{k}\times
   \begin{cases}
  l(2a,k)l(\epsilon_{1}+\epsilon_{2}+2a,k)\quad\qquad\:\,\text{if}\quad k>0,\\
  l(-2a,-k)l(\epsilon_{1}+\epsilon_{2}-2a,-k)\quad\text{if}\quad k<0,
   \end{cases}
\end{equation}
where
\begin{equation*}
 l(x,n)=\prod_{\substack{i,j\geq1,\;i+j\leq2n\\i+j\equiv0\mod 2}}\hspace*{-10pt}(x+(i-1)\epsilon_{1}+(j-1)\epsilon_{2}).
\end{equation*}
Two factors $Z_{\textsf{vec}}^{(2)}$ in \eqref{Zvec-def-2} arise from the points $p_1, p_2 \in X_2$ invariant under the torus action. The factor $l_{\vec{k}}$ arises from the exceptional divisor. We will call this factor as blow-up factor.

The instanton part of the Nekrasov partition function for the pure $U(2)$ gauge theory on $X_{2}$ can be written as \cite{Bonelli:2011jx}
\begin{equation}\label{BMT-formula}
   Z_{\text{pure}}^{(2,X_{2})}(\vec{a},\epsilon_{1},\epsilon_{2}|\Lambda)=\sum_{k\in\mathbb{Z}}
   \frac{\Lambda^{2k^{2}}}{l_{\vec{k}}(\vec{a}|\epsilon_{1},\epsilon_{2})}
   Z_{\text{pure}}^{(2)}(\vec{a}+\epsilon_{1}\vec{k},2\epsilon_{1},\epsilon_{2}-\epsilon_{1}|\Lambda)
   Z_{\text{pure}}^{(2)}(\vec{a}+\epsilon_{2}\vec{k},\epsilon_{1}-\epsilon_{2},2\epsilon_{2}|\Lambda),
\end{equation}
where $Z_{\text{pure}}^{(2)}(\vec{a},\epsilon_{1},\epsilon_{2}|\Lambda)$ is given by \eqref{Zpure-def}. Equations \eqref{Zvec-def-2} and  \eqref{BMT-formula} give some hint about the structure of  the basis of states in this case. Namely, the r.h.s. of \eqref{BMT-formula} is expressed in terms of two partition functions (corresponding to the case $p=1$, $r=2$ from our scheme) with parameters
\begin{equation}
  \begin{aligned}
    &\epsilon_{1}^{\scriptscriptstyle{(1)}}=2\epsilon_{1},&\qquad
    &\epsilon_{2}^{\scriptscriptstyle{(1)}}=\epsilon_{2}-\epsilon_{1},\\
    &\epsilon_{1}^{\scriptscriptstyle{(2)}}=\epsilon_{1}-\epsilon_{2},&\qquad
    &\epsilon_{2}^{\scriptscriptstyle{(2)}}=2\epsilon_{2}.
  \end{aligned}
\end{equation}
We note that if we define CFT parameters $b^{\scriptscriptstyle{(\sigma)}}$ by
\begin{equation*}
   (b^{\scriptscriptstyle{(\sigma)}})^{2}=\frac{\epsilon_{1}^{\scriptscriptstyle{(\sigma)}}}{\epsilon_{2}^{\scriptscriptstyle{(\sigma)}}},
\end{equation*}
then they are subject to the relation
\begin{equation}\label{bb-relat-1}
   (b^{\scriptscriptstyle{(1)}})^{2}+(b^{\scriptscriptstyle{(2)}})^{-2}=-2.
\end{equation}
One can propose that similar relation should hold in the CFT terms too. Namely, in algebraic language we expect that in the algebra $\mathcal{H}\oplus\mathcal{H}\oplus\mathcal{F}\oplus \textsf{NSR}$ there are two commuting subalgebras $\mathcal{H}\oplus\mathsf{Vir}$ with the parameters $b^{\scriptscriptstyle{(1)}}$ and $b^{\scriptscriptstyle{(2)}}$ satisfying \eqref{bb-relat-1}. In the next subsection we give explicit construction of these two subalgebras.


\subsection{Algebraic setup}
As was claimed above  this case corresponds to the algebra $\mathcal{A}=\mathcal{H}\oplus\mathcal{H}\oplus\mathcal{F}\oplus \textsf{NSR}$. Let us first introduce the notations. The commutation relations of the Neveu-Schwarz-Ramond algebra are known to be
\begin{equation}\label{NS-comm-relat}
\begin{aligned}
&[L_{n},L_{m}]=(n-m)L_{n+m}+\frac{c_{\textsf{\tiny{NSR}}}}{8}(n^{3}-n)\delta_{n+m},\\
&\{G_{r},G_{s}\}= 2L_{r+s}+\frac{1}{2}c_{\textsf{\tiny{NSR}}}(r^{2}-\frac{1}{4})\delta_{r+s,0},\\
&[L_{n},G_{r}]=\left(\frac{1}{2}n-r\right)G_{n+r}.
\end{aligned}
\end{equation}
The central charge $c_{\scriptscriptstyle{\textsf{NSR}}}$ is parameterized as follows
\begin{equation}
c_{\textsf{\tiny{NSR}}}=1+2Q^{2}, \quad Q=b+\frac{1}{b}.
\end{equation}
The indexes $r$ and $s$ in \eqref{NS-comm-relat} are either  integer (the Ramond sector), or half and odd integer (the Neveu-Schwarz sector). Below we will  consider the Neveu-Schwarz  sector. The highest weight representation in this case is defined by the vacuum state $|P\rangle_{\scriptscriptstyle{\textsf{NS}}}$
\begin{equation}\label{NS-vac}
    L_{n}|P\rangle_{\scriptscriptstyle{\textsf{NS}}}=G_{r}|P\rangle_{\scriptscriptstyle{\textsf{NS}}}=0\quad\text{for}\quad n,r>0,\qquad
    L_{0}|P\rangle_{\scriptscriptstyle{\textsf{NS}}}=\Delta_{\textsf{\tiny{NS}}}(Q/2+P,b)|P\rangle_{\scriptscriptstyle{\textsf{NS}}},
\end{equation}
where
\begin{equation}\label{Delta-NS}
   \Delta_{\textsf{\tiny{NS}}}(\alpha,b)=\frac{1}{2}\alpha(Q-\alpha).
\end{equation}
\subsubsection{Two commutative Virasoro algebras}
We will extend our algebra  multiplying it by two additional Heisenberg algebras $\mathcal{H}$  and one fermion algebra $\mathcal{F}$. Let us first multiply the \textsf{NSR} algebra by additional  fermion (in the Neveu-Schwarz sector)
\begin{equation}\label{free-fermion}
\{f_{r},f_{s}\}=\delta_{r+s,0},\quad
 r,s\in\mathbb{Z}+\frac{1}{2}
\end{equation}
and also we assume that it anticommutes with generators $G_{r}$
\begin{equation}
   \{G_{r},f_{s}\}=0.
\end{equation}
It was pointed out in  \cite{Crnkovic:1989gy,Crnkovic:1989ug,Lashkevich:1992sb} that there exists a non-trivial embedding of two commuting Virasoro algebras in $\mathcal{F}\oplus\mathsf{NSR}$ which will be an essential point of our construction\footnote{The possibility of using  of the construction \cite{Crnkovic:1989gy,Crnkovic:1989ug,Lashkevich:1992sb} in this context was also suggested by Wyllard in \cite{Wyllard:2011mn}.}.
Following \cite{Crnkovic:1989gy,Crnkovic:1989ug,Lashkevich:1992sb} we can notice that the combinations
\begin{equation}\label{2-Vir}
\begin{aligned}
&L_{n}^{\scriptscriptstyle{(1)}}= \frac{1}{1-b^{2}}L_{n} -\frac{1+2b^{2}}{2(1-b^{2})}\sum_{r=-\infty}^{\infty}r:f_{n-r}f_{r}:+\frac{b}{1-b^{2}}\sum_{r=-\infty}^{\infty}f_{n-r}G_{r} ,\\
&L_{n}^{\scriptscriptstyle{(2)}}= \frac{1}{1-b^{-2}}L_{n} -\frac{1+2b^{-2}}{2(1-b^{-2})}\sum_{r=-\infty}^{\infty}r:f_{n-r}f_{r}:+\frac{b^{-1}}{1-b^{-2}}\sum_{r=-\infty}^{\infty}f_{n-r}G_{r},
\end{aligned}
\end{equation}
commute with  each other and satisfy the Virasoro commutation relations i.e.
\begin{equation}\label{two-Virasoro}
   \begin{gathered}
     [L_{n}^{\scriptscriptstyle{(1)}},L_{m}^{\scriptscriptstyle{(2)}}]=0,\\
     [L_{n}^{\scriptscriptstyle{(\sigma)}},L_{m}^{\scriptscriptstyle{(\sigma)}}]=(n-m)L_{n+m}^{\scriptscriptstyle{(\sigma)}}+\frac{c^{\scriptscriptstyle{(\sigma)}}}{12}(n^{3}-n)\,\delta_{n+m,0},
   \end{gathered}
\end{equation}
with
\addtocounter{equation}{-1}
\begin{subequations}
\begin{equation}\label{newbQ}
   c^{\scriptscriptstyle{(\sigma)}}=1+6Q^{(\sigma)}\,^{2}, \quad
   Q^{\scriptscriptstyle{(\sigma)}}=b^{(\sigma)}+1/b^{(\sigma)}\quad\text{and}\quad
   b^{\scriptscriptstyle{(1)}}  =\frac{2b}{\sqrt{2-2b^{2}}},\quad
   (b^{\scriptscriptstyle{(2)}})^{-1}=\frac{2b^{-1}}{\sqrt{2-2b^{-2}}}.
\end{equation}
\end{subequations}
We note that the parameters $b^{\scriptscriptstyle{(1)}}$ and $b^{\scriptscriptstyle{(2)}}$ satisfy the relation \eqref{bb-relat-1}.

Consider the highest weight representation  $\pi_{\scriptscriptstyle{\mathcal{F}\oplus\mathsf{NSR}}}=\pi_{\scriptscriptstyle{\mathcal{F}}}\otimes\pi_{\scriptscriptstyle{\mathsf{NSR}}}$ of the algebra $\mathcal{F}\oplus\mathsf{NSR}$. In other words we extend the definition of the highest weight vector  \eqref{NS-vac} by demanding that
\begin{equation*}
    f_{r}|P\rangle_{\scriptscriptstyle{\textsf{NS}}}=0,\qquad\text{for}\quad r>0.
\end{equation*}
For general values of the momenta $P$ the highest weight  representation $\pi_{\scriptscriptstyle\mathcal{F}\oplus\mathsf{NSR}}$  is irreducible. Its character is given by
\begin{equation}\label{Char-NSfer}
   \chi_{\scriptscriptstyle\mathcal{F}\oplus\mathsf{NSR}}(q)=\chi_{\mathcal{F}}(q)^{2}\chi_{\mathcal{B}}(q),
\end{equation}
where
\begin{equation*}
   \chi_{\mathcal{F}}(q)=\prod_{k>0}(1+q^{k-\frac{1}{2}}),\qquad
   \chi_{\mathcal{B}}(q)=\prod_{k>0}\frac{1}{(1-q^{k})}
\end{equation*}
are  fermionic and bosonic  characters\footnote{Usually, the character which  is defined as $\textrm{Tr}\,q^{L_{0}}\bigl|_{\pi_{\Delta}}$ is proportional to $q^{\Delta}$. We erased these factors for simplicity.}.

\begin{figure}
\psfrag{k}{$k$}
\psfrag{a1}{$\scriptstyle -2$}
\psfrag{a2}{$\scriptstyle -1$}
\psfrag{a3}{$\scriptstyle 0$}
\psfrag{a4}{$\scriptstyle 1$}
\psfrag{a5}{$\scriptstyle 2$}
\psfrag{d}{$\frac{1}{2}$}
\psfrag{l}{$L_{0}$}
\psfrag{b1}{$\scriptstyle 0$}
\psfrag{b2}{$\scriptstyle \frac{1}{2}$}
\psfrag{b3}{$\scriptstyle 1$}
\psfrag{b4}{$\scriptstyle \frac{3}{2}$}
\psfrag{b5}{$\scriptstyle 2$}
\psfrag{b6}{$\scriptstyle \frac{5}{2}$}
	\centering
	\includegraphics[width=.6\textwidth]{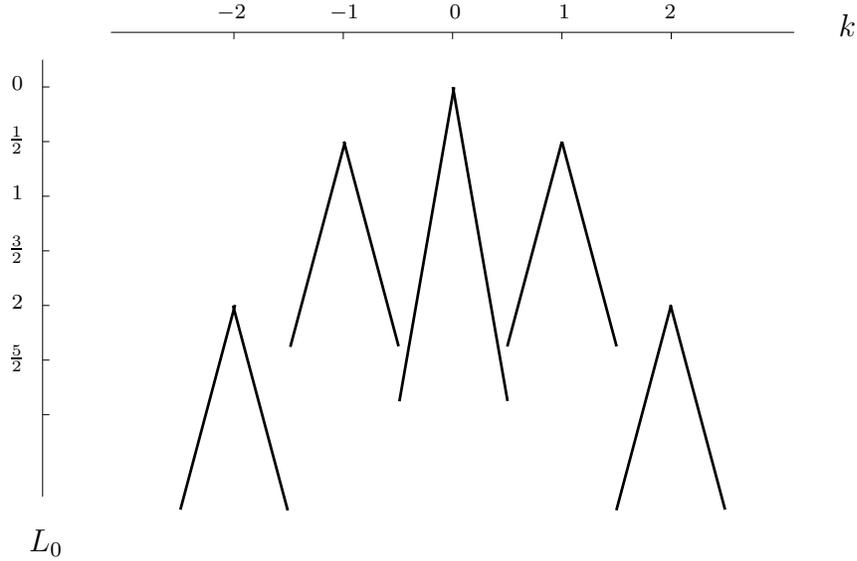}
        \caption{Decomposition of an irreducible representation of the algebra $\mathcal{F}\oplus\mathsf{NSR}$ into direct sum of representations of the algebra $\mathsf{Vir}\oplus\mathsf{Vir}$. Each interior angle corresponds to Verma module $\pi^{k}_{\scriptscriptstyle\mathsf{Vir}\oplus \mathsf{Vir}}$ over the algebra $\mathsf{Vir}\oplus\mathsf{Vir}$ whose conformal dimension is shifted by $k^{2}/2$ as in  \eqref{Delta-shifted}.}
	\label{NS-decompos-pic}
\end{figure}

We see from \eqref{2-Vir} that there is a natural action of the two Virasoro algebras in the representation $\pi_{\scriptscriptstyle\mathcal{F}\oplus\mathsf{NSR}}$. As a representation of  $\mathsf{Vir}\oplus\mathsf{Vir}$ it is no longer irreducible and for general values of the momenta $P$ can be decomposed into direct sum of the Verma modules $\pi_{\scriptscriptstyle\mathsf{Vir}\oplus \mathsf{Vir}}$ over the algebra $\mathsf{Vir}\oplus\mathsf{Vir}$.
The character of any of $\pi_{\scriptscriptstyle\mathsf{Vir}\oplus \mathsf{Vir}}$ is given by
\begin{equation}\label{Char-VirVir}
   \chi_{\scriptscriptstyle\mathsf{Vir}\oplus \mathsf{Vir}}(q)=\chi_{\mathcal{B}}(q)^{2}.
\end{equation}
Using the consequence of the Jabobi triple product identity
\begin{equation*}
    \prod_{k>0}(1+q^{k-\frac{1}{2}})^{2}(1-q^{k})=\sum_{k\in\mathbb{Z}}q^{\frac{k^{2}}{2}}=1+2q^{\frac{1}{2}}+2q^{2}+2q^{\frac{9}{2}}+\dots
\end{equation*}
we see that
\begin{equation}\label{char-decomposition}
   \chi_{\scriptscriptstyle\mathcal{F}\oplus\mathsf{NSR}}(q)=
   \sum_{k\in\mathbb{Z}}q^{\frac{k^{2}}{2}}\chi_{\scriptscriptstyle\mathsf{Vir}\oplus \mathsf{Vir}}(q),
\end{equation}
which implies the decomposition (see fig. \ref{NS-decompos-pic})
\begin{equation}\label{NS-decompos}
   \pi_{\scriptscriptstyle\mathcal{F}\oplus\mathsf{NSR}}=\bigoplus_{k\in\mathbb{Z}} \pi^{k}_{\scriptscriptstyle\mathsf{Vir}\oplus \mathsf{Vir}},
\end{equation}
where $\pi^{k}_{\scriptscriptstyle\mathsf{Vir}\oplus \mathsf{Vir}}$ is the Verma module of $\mathsf{Vir}\oplus\mathsf{Vir}$ with the highest weight $|P,k\rangle$. The highest weight state $|P,k\rangle$ is defined as
\begin{equation}\label{high-weight-def}
\begin{gathered}
   L_{n}^{\scriptscriptstyle{(1)}}|P,k\rangle=L_{n}^{\scriptscriptstyle{(2)}}|P,k\rangle=0\qquad\text{for}\qquad n>0,\\
   L_{0}^{\scriptscriptstyle{(1)}}|P,k\rangle=\Delta^{\scriptscriptstyle{(1)}}(P,k)|P,k\rangle,\qquad
   L_{0}^{\scriptscriptstyle{(2)}}|P,k\rangle=\Delta^{\scriptscriptstyle{(2)}}(P,k)|P,k\rangle,
\end{gathered}
\end{equation}
where the conformal dimensions $\Delta^{\scriptscriptstyle{(1)}}(P,k)$ and $\Delta^{\scriptscriptstyle{(2)}}(P,k)$ satisfy the relation
\begin{equation}\label{Delta-shifted}
    \Delta^{\scriptscriptstyle{(1)}}(P,k)+\Delta^{\scriptscriptstyle{(2)}}(P,k)=\Delta_{\scriptscriptstyle{\textsf{NS}}}(Q/2+P,b)+\frac{k^{2}}{2}.
\end{equation}
Equation \eqref{Delta-shifted} follows from the relation
\begin{equation*}
  L_{0}^{\scriptscriptstyle{(1)}}+L_{0}^{\scriptscriptstyle{(2)}}=L_{0}+L_{0}^{\textrm{f}},
\end{equation*}
where $L_{0}^{\textrm{f}}$ is the zeroth component of the stress-energy tensor for the free-fermion
\begin{equation*}
  L_{0}^{\textrm{f}}=\sum_{r=1/2}^{\infty}r f_{-r}f_{r}.
\end{equation*}

In order to construct the highest weight states $|P,k\rangle$ in more explicit terms and to compute the conformal dimensions $\Delta^{\scriptscriptstyle{(1)}}(P,k)$ and $\Delta^{\scriptscriptstyle{(2)}}(P,k)$ we consider free-field representation for the \textsf{NSR} algebra. There exist two alternative free-field representations (corresponding to the choice of sign in front of operator $\mathcal{P}$)
\begin{equation}\label{NS-bosonization}
\begin{aligned}
&L_{n}= \frac{1}{2}\sum_{k\neq 0,n}c_{k}c_{n-k}+\frac{1}{2}\sum_{r}(r-\frac{n}{2})\psi_{n-r}\psi_{r}+\frac{i}{2}(Qn\mp 2\mathcal{P})c_{n},\\
&L_{0}=\sum_{k>0}c_{-k}c_{k}+\sum_{r>0}r\psi_{-r}\psi_{r}+\frac{1}{2}\big(\frac{Q^{2}}{4}-\mathcal{P}^{2}\big),\\
&G_{r}= \sum_{n\neq 0}c_{n}\psi_{r-n}+i(Qr\mp \mathcal{P})\psi_{r},\qquad
\mathcal{P}|P\rangle_{\scriptscriptstyle{\textsf{NS}}}=P|P\rangle_{\scriptscriptstyle{\textsf{NS}}},
\end{aligned}
\end{equation}
where the operator of zero mode $\mathcal{P}$, bosonic components $c_{n}$ and fermionic components $\psi_{r}$ satisfy commutation relations
\begin{equation}\label{NS-bosonization-comm-relat}
\begin{gathered}
[c_{n},c_{m}]= n\,\delta_{n+m,0},\quad
\{\psi_{r},\psi_{s}\} =\delta_{r+s,0},\\
[\mathcal{P},c_{n}]=[\mathcal{P},\psi_{r}]=0.
\end{gathered}
\end{equation}
It is convenient to introduce the combinations
\begin{equation*}
   \chi_{r}=f_{r}-i\psi_{r},
\end{equation*}
then one can show that the state
\begin{equation}\label{high-weight-def2}
   |P,k\rangle=\Omega_{k}(P)\,\chi_{-\frac{1}{2}}\chi_{-\frac{3}{2}}\dots\chi_{-\frac{2|k|-1}{2}}|\text{\textrm{vac}}\rangle,
\end{equation}
is the highest weight vector, i.e. it satisfies the conditions \eqref{high-weight-def}
and $|\text{\textrm{vac}}\rangle$ is the vacua state defined by
\begin{equation*}
   c_{n}|\text{\textrm{vac}}\rangle=\psi_{r}|\text{\textrm{vac}}\rangle=f_{r}|\text{\textrm{vac}}\rangle=0,\quad\text{for}\quad n,r>0.
\end{equation*}
Last statement can be derived using the relations
\begin{equation}
 \begin{aligned}
   &[L_{n}^{\scriptscriptstyle{(1)}}+L_{n}^{\scriptscriptstyle{(2)}},\chi_{r}]=-\left(\frac{n}{2}+r\right)\chi_{r+n},\\
   &[bL_{n}^{\scriptscriptstyle{(1)}}+b^{-1}L_{n}^{\scriptscriptstyle{(2)}},\chi_{r}]=-\left((n+r)Q\mp\mathcal{P}\right)\chi_{r+n}+
   i\sum_{m \neq 0}c_{m}\chi_{r+n-m}.
 \end{aligned}
\end{equation}
The choice of sign in front of the operator of the zero mode $\mathcal{P}$ in \eqref{NS-bosonization} corresponds to $k>0$ or $k<0$ in  \eqref{high-weight-def2}. Choosing ``$\mp$'' in  \eqref{NS-bosonization} we define two different sets of generators $c_{k}$ and $\psi_{r}$. Similarly to the bosonic case they are related by some unitary transform (in particular if $Q=0$ they just differ by a sign).

Using \eqref{high-weight-def2} one can compute
\begin{equation}
  \Delta^{\scriptscriptstyle{(1)}}(P,k)=\frac{(Q^{\scriptscriptstyle{(1)}})^{2}}{4}-\left(P^{\scriptscriptstyle{(1)}}
  +\frac{kb^{\scriptscriptstyle{(1)}}}{2}\right)^{2},\quad
  \Delta^{\scriptscriptstyle{(2)}}(P,k)=\frac{(Q^{\scriptscriptstyle{(2)}})^{2}}{4}-\left(P^{\scriptscriptstyle{(2)}}
  +\frac{k}{2b^{\scriptscriptstyle{(2)}}}\right)^{2},
\end{equation}
where parameters $b^{\scriptscriptstyle{(\sigma)}}$ and $Q^{\scriptscriptstyle{(\sigma)}}$ are given by \eqref{newbQ} and
\begin{equation}\label{newP}
   P^{\scriptscriptstyle{(1)}}=\frac{P}{\sqrt{2-2b^{2}}}\quad\text{and\quad}P^{\scriptscriptstyle{(2)}}=\frac{P}{\sqrt{2-2b^{-2}}}.
\end{equation}
One can also define the state $\langle k',P'|$ conjugated to \eqref{high-weight-def2}
\begin{equation}\label{high-weight-def2-dual}
   \langle k',P'|=\Omega_{k'}(P')\langle\text{\textrm{vac}}|\chi_{\frac{2|k'|-1}{2}}\dots\chi_{\frac{1}{2}}.
\end{equation}
This choice is consistent with the following conjugation $f_{r}^{+}=-f_{-r}$.
We  chose the normalization factors $\Omega_{k}(P)$ in \eqref{high-weight-def2} and \eqref{high-weight-def2-dual} such that
\begin{equation}
   |P,k\rangle=\left(\bigl(G_{-\frac{1}{2}}\bigr)^{k^{2}}+\dots\right)|P\rangle,\qquad
   \langle k',P'|=\langle P'|\left(\bigl(G_{\frac{1}{2}}\bigr)^{k'^{2}}+\dots\right),
\end{equation}
where omitted terms have smaller degree in $G$. One can find that
\begin{equation}
   \Omega_{k}(P)=\frac{1}{2}\,\prod_{m+n\leq 2|k|}(2P+mb+nb^{-1}).
\end{equation}
This normalization is standard in CFT and from the other side it coincides with geometrical normalization. The norm of the state $|P,k\rangle$ equals to the determinant of the vector field\footnote{Note that states $|P,k\rangle$ and $\langle k',P'|$ cannot be represented in form \eqref{high-weight-def} and \eqref{high-weight-def2-dual} simultaneously.}
\begin{equation}
  \langle k,P|P,k\rangle=\det v\Bigl|_{p_{\scriptscriptstyle{(\varnothing,\varnothing),(\varnothing,\varnothing)},k}}
\end{equation}
and coincides with the factor \eqref{l-vec}.
\subsubsection{Construction of the basis}
Now we can multiply our algebra $\mathcal{F}\oplus\mathsf{NSR}$ by two additional Heisenberg algebras $\mathcal{H}\oplus\mathcal{H}$ with generators $h_{n}$ and $w_{n}$
\begin{equation}
 [h_{n},h_{m}]=[w_{n},w_{m}]=n\delta_{n+m,0},\qquad
 [h_{n},w_{m}]=0.
\end{equation}
The sets of bosons $w_{n}$ and $h_{n}$ have different nature. In particular, the bosons $w_{n}$ are analogous to the bosons $\mathtt{a}_{n}$ and $a_{n}$ considered in section \ref{AFLT} and enter into vertex operators in non-symmetric way (see e.g. \eqref{super-primary}--\eqref{vertex-super} and compare it to \eqref{vertex-CO} and \eqref{vertex}). Contrary, the bosons $h_{n}$ always enter in vertex operators in a symmetric way (see \eqref{super-primary-m}). From the point of view of scheme \eqref{scheme-1} the  bosons $w_{n}$ correspond to the factor $\mathcal{H}$ in $\mathcal{H}\oplus\widehat{\mathfrak{sl}}(2)_{2}\oplus\mathsf{NSR}$, while the bosons $h_{n}$ belong to the free-field representation for  $\widehat{\mathfrak{sl}}(2)_{2}$ algebra.

We define also another set of generators
\begin{equation}\label{rotated-bosons}
   a_{n}^{\scriptscriptstyle{(1)}}=\frac{1}{\sqrt{2-2b^{2}}}(w_{n}-ibh_{n}),\qquad
   a_{n}^{\scriptscriptstyle{(2)}}=\frac{1}{\sqrt{2-2b^{-2}}}(w_{n}-ib^{-1}h_{n}),
\end{equation}
such that
\begin{equation}\label{rotated-bosons-comm-relat}
   [a_{n}^{\scriptscriptstyle{(\sigma)}},a_{m}^{\scriptscriptstyle{(\rho)}}]=\frac{n}{2}\delta_{n+m,0}\,\delta_{\sigma,\rho},\qquad
   \sigma,\rho=1,2.
\end{equation}
Thus in the algebra $\mathcal{H}\oplus\mathcal{H}\oplus\mathcal{F}\oplus \textsf{NSR}$ we have two subalgebras $\mathcal{H}\oplus\mathsf{Vir}$ with generators $a_{n}^{\scriptscriptstyle{(\sigma)}}$ and $L_{n}^{\scriptscriptstyle{(\sigma)}}$ for $\sigma=1,2$ which satisfy \eqref{two-Virasoro}, \eqref{rotated-bosons-comm-relat} and obvious relations
\begin{equation*}
   [L_{n}^{\scriptscriptstyle{(\sigma)}},a_{m}^{\scriptscriptstyle{(\rho)}}]=0.
\end{equation*}
We note that bosons $a_{n}^{\scriptscriptstyle{(1)}}$ and $a_{n}^{\scriptscriptstyle{(2)}}$ enter in our construction in a completely symmetric way (together with the symmetry $b\rightarrow1/b$). For each of these subalgebras we can define integrable system  \eqref{I1I2}:
\begin{equation}\label{I1I2-new}
  \begin{aligned}
     &\mathbf{I}_{1}^{\scriptscriptstyle{(\sigma)}}=L_{0}^{\scriptscriptstyle{(\sigma)}}+2\sum_{k>0}a_{-k}^{\scriptscriptstyle{(\sigma)}}a_{k}^{\scriptscriptstyle{(\sigma)}},\\
     &\mathbf{I}_{2}^{\scriptscriptstyle{(\sigma)}}= \sum_{k\neq0}a_{-k}^{\scriptscriptstyle{(\sigma)}}L_{k}^{\scriptscriptstyle{(\sigma)}}+2iQ\sum_{k>0}^{\infty}ka_{-k}^{\scriptscriptstyle{(\sigma)}}a_{k}^{\scriptscriptstyle{(\sigma)}}+
     \frac{1}{3}\sum_{i+j+k=0}a_{i}^{\scriptscriptstyle{(\sigma)}}a_{j}^{\scriptscriptstyle{(\sigma)}}a_{k}^{\scriptscriptstyle{(\sigma)}}.
  \end{aligned}
\end{equation}
The eigenvectors for this integrable system can be easily found. At first, we redefine the highest weight states \eqref{high-weight-def}  by demanding that
\begin{equation*}
  h_{n}|P,k\rangle=w_{n}|P,k\rangle=0\quad\text{for}\quad n>0.
\end{equation*}
Then the eigenvectors can be written in the form
\begin{equation}\label{new-eigenfunctions}
   |P,k\rangle_{\scriptscriptstyle{\vec{Y}}^{(1)},\scriptscriptstyle{\vec{Y}}^{(2)}}\overset{\text{def}}{=}
   X_{\scriptscriptstyle{\vec{Y}}^{(1)}}
   \Bigl(P^{\scriptscriptstyle{(1)}}+\frac{kb^{\scriptscriptstyle{(1)}}}{2},b^{\scriptscriptstyle{(1)}}  \Bigr)
   X_{\scriptscriptstyle{\vec{Y}}^{(2)}}
   \Bigl(P^{\scriptscriptstyle{(2)}}+\frac{k}{2b^{\scriptscriptstyle{(2)}}},b^{\scriptscriptstyle{(2)}}  \Bigr)|P,k\rangle,
\end{equation}
where $\vec{Y}^{\scriptscriptstyle{(1)}}$ and $\vec{Y}^{\scriptscriptstyle{(2)}}$ are two pairs of the Young diagrams and parameters $b^{(\sigma)}$ and $P^{(\sigma)}$ are given by \eqref{newbQ} and \eqref{newP}. Operators $X_{\scriptscriptstyle{\vec{Y}^{(\sigma)}}}(P^{\scriptscriptstyle{(\sigma)}},b^{\scriptscriptstyle{(\sigma)}})$ in \eqref{new-eigenfunctions} are given by \eqref{X-def} and  consists of generators $L_{-n}^{\scriptscriptstyle{(\sigma)}}$ and $a_{-n}^{\scriptscriptstyle{(\sigma)}}$.

We claim that the basis \eqref{new-eigenfunctions} factorizes certain primary operators analogous to \eqref{primary}. It is remarkable that  compared to the case $p=1$ we have infinitely many of them
\begin{equation}\label{super-primary-k}
   \mathbb{V}_{\alpha}^{(m)}\qquad m\in\mathbb{Z},
\end{equation}
which corresponds  to the highest weight states $|P,m\rangle$ due to the operator--state correspondence. Only the field $V_{\alpha}^{(0)}$ corresponds to the primary field of the \textsf{NSR} algebra, the rest correspond to descendant fields  with the conformal dimensions under the ``total'' stress-energy tensor $T(z)+T^{\textrm{f}}(z)$
\begin{equation*}
   \Delta_{\scriptscriptstyle{\textsf{NS}}}(\alpha)+\frac{m^{2}}{2},
\end{equation*}
where $T^{\textrm{f}}(z)$ is the stress-energy tensor for the Majorana fermion $f_{r}$. The first few examples of the fields $V_{\alpha}^{(m)}$ can be easily calculated:
\begin{equation}\label{super-primary}
\begin{aligned}
   &\mathbb{V}_{\alpha}^{\scriptscriptstyle{(0)}}(z)=\Phi_{\alpha}^{\scriptscriptstyle{\textsf{NS}}}(z)\cdot \mathcal{W}_{\alpha}(z),\\
   &\mathbb{V}_{\alpha}^{\scriptscriptstyle{(1)}}(z)=\bigl(\alpha f(z)\Phi_{\alpha}^{\scriptscriptstyle{\textsf{NS}}}(z)+
   \Psi_{\alpha}^{\scriptscriptstyle{\textsf{NS}}}(z)\bigr)\,e^{i\phi(z)}\,\mathcal{W}_{\alpha}(z),\\
   &\mathbb{V}_{\alpha}^{\scriptscriptstyle{(-1)}}(z)=\bigl((Q-\alpha) f(z)\Phi_{\alpha}^{\scriptscriptstyle{\textsf{NS}}}(z)+
   \Psi_{\alpha}^{\scriptscriptstyle{\textsf{NS}}}(z)\bigr)\,e^{-i\phi(z)}\,\mathcal{W}_{\alpha}(z),
\end{aligned}
\end{equation}
where $\Phi_{\alpha}^{\scriptscriptstyle{\textsf{NS}}}$ is the primary field of the \textsf{NSR} algebra with conformal
dimension $\Delta(\alpha)=\frac{1}{2}\alpha(Q-\alpha)$, $\Psi_{\alpha}^{\scriptscriptstyle{\textsf{NS}}}$ its super partner with the dimension $\Delta(\alpha)+1/2$,
\begin{equation*}
f(z)=\sum_{r}f_{r}z^{r+1/2},\qquad \phi(z)=i\sum_{n\neq0}\frac{h_{n}}{n}z^{-n}
\end{equation*}
and $\mathcal{W}_{\alpha}$ is a free exponential
\begin{equation}\label{vertex-super}
\mathcal{W}_{\alpha}= e^{(\alpha-Q)\varphi_{-}}e^{\alpha\varphi_{+}},
\end{equation}
with $\varphi_{+}=i\sum_{n>0}\frac{w_{n}}{n}z^{-n}$ and $\varphi_{-}(z)=i\sum_{n<0}\frac{w_{n}}{n}z^{-n}$. For general $m$ the field $\mathbb{V}_{\alpha}^{(m)}$ has a form
\begin{equation}\label{super-primary-m}
   \mathbb{V}_{\alpha}^{(m)}=D^{m}[\Phi_{\alpha}^{\scriptscriptstyle{\textsf{NS}}}(z),f(z)]\,
   e^{im\phi(z)}\,\mathcal{W}_{\alpha}(z),
\end{equation}
where $D^{m}[\Phi_{\alpha}^{\scriptscriptstyle{\textsf{NS}}}(z),f(z)]$ is some descendant field on a level $m^{2}/2$.\footnote{Geometrical definition of the vertex operator in \cite{Carlsson:2008fk} (for the case of Hilbert schemes) depends on the line bundle on the surface. It is natural to expect that the vertex operator $\mathbb{V}_{\alpha}^{(m)}$ corresponds to the line bundle $\mathcal{O}(mC)$ on the surface $X_2$}

The commutation relations of the primary fields $\Phi_{\alpha}^{\scriptscriptstyle{\textsf{NS}}}$, $\Psi_{\alpha}^{\scriptscriptstyle{\textsf{NS}}}$ and $\mathcal{W}_{\alpha}$ with generators $L_{n}$, $a_{n}$, $w_{n}$, $G_{r}$ and $f_{r}$ can be summarized as
\begin{equation}
\begin{aligned}
&[L_{n},\Phi_{\alpha}^{\scriptscriptstyle{\textsf{NS}}}] =
(z^{n+1}\partial_{z}+(n+1)\Delta(\alpha)z^{n})\Phi_{\alpha}^{\scriptscriptstyle{\textsf{NS}}},\\
& [L_{n},\Psi_{\alpha}^{\scriptscriptstyle{\textsf{NS}}}] =
(z^{n+1}\partial_{z}+(n+1)(\Delta(\alpha)+1/2)z^{n})\Psi_{\alpha}^{\scriptscriptstyle{\textsf{NS}}},\\
&[G_{r},\Phi_{\alpha}^{\scriptscriptstyle{\textsf{NS}}}] =
z^{r+1/2}\Psi_{\alpha}^{\scriptscriptstyle{\textsf{NS}}},\\
&\{G_{r},\Psi_{\alpha}^{\scriptscriptstyle{\textsf{NS}}}\}=
(z^{r+1/2}\partial_{z}+(2r+1)\Delta(\alpha)z^{r-1/2})\Phi_{\alpha}^{\scriptscriptstyle{\textsf{NS}}},\\
& [w_{n},\mathcal{W}_{\alpha}(z)]=-i\alpha z^{n}\mathcal{W}_{\alpha}, \quad\qquad \textrm{for}\; n<0,\\
& [w_{n},\mathcal{W}_{\alpha}(z)]=i(Q-\alpha)z^{n}\mathcal{W}_{\alpha}, \quad \textrm{for}\; n>0.
\end{aligned}
\end{equation}
Let us  consider the matrix elements
\begin{equation}
   \mathfrak{F}(\alpha,m|P',k',\vec{W}^{\scriptscriptstyle{(1)}},\vec{W}^{\scriptscriptstyle{(2)}};
   P,k,\vec{Y}^{\scriptscriptstyle{(1)}},\vec{Y}^{\scriptscriptstyle{(2)}})\overset{\text{def}}{=}\frac
   {_{\scriptscriptstyle{\vec{W}^{(1)}},\scriptscriptstyle{\vec{W}^{(2)}}}\langle k',P'|
   \mathbb{V}_{\alpha}^{(m)}|P,k\rangle_{\scriptscriptstyle{\vec{Y}}^{(1)},\scriptscriptstyle{\vec{Y}}^{(2)}}}
   {\langle k',P'|\mathbb{V}_{\alpha}^{(m)}|P,k\rangle}.
\end{equation}
\begin{Proposition}\label{main-prop}
We propose that
\begin{multline}\label{main-prop-equality}
\mathfrak{F}(\alpha,m|P',k',\vec{W}^{\scriptscriptstyle{(1)}},\vec{W}^{\scriptscriptstyle{(2)}};
   P,k,\vec{Y}^{\scriptscriptstyle{(1)}},\vec{Y}^{\scriptscriptstyle{(2)}})=
  \mathbb{F}\Bigl(\alpha^{\scriptscriptstyle{(1)}}+\frac{mb^{\scriptscriptstyle{(1)}}}{2},b^{\scriptscriptstyle{(1)}}  \Bigl|P'_{1}+
  \frac{k'b^{\scriptscriptstyle{(1)}}  }{2},\vec{W}^{\scriptscriptstyle{(1)}},P_{1}+\frac{kb^{\scriptscriptstyle{(1)}}  }{2},
  \vec{Y}^{\scriptscriptstyle{(1)}}\Bigr)
  \times\\\times
  \mathbb{F}\Bigl(\alpha^{\scriptscriptstyle{(2)}}+\frac{m}{2b^{\scriptscriptstyle{(2)}}},b^{\scriptscriptstyle{(2)}}\Bigr|
  P'_{2}+\frac{k'}{2b^{\scriptscriptstyle{(2)}}},\vec{W}^{\scriptscriptstyle{(2)}},
  P_{2}+\frac{k}{2b^{\scriptscriptstyle{(2)}}},\vec{Y}^{\scriptscriptstyle{(2)}}\Bigr),
\end{multline}
where
\begin{equation*}
   \alpha^{\scriptscriptstyle{(1)}}=\frac{\alpha}{\sqrt{2-2b^{2}}},\quad
   \alpha^{\scriptscriptstyle{(2)}}=\frac{\alpha}{\sqrt{2-2b^{-2}}};
\end{equation*}
and parameters $b_{j}$ and $P_{j}$ are given by \eqref{newbQ} and \eqref{newP} and function $\mathbb{F}$ by \eqref{Zbif-def}--\eqref{Zbif-def-2}.
\end{Proposition}
We note that Proposition \ref{main-prop} suggests the following identification
\begin{equation}\label{strange-relation}
   \mathbb{V}_{\alpha}^{(m)}(z)=V_{\alpha^{\scriptscriptstyle{(1)}}+mb^{\scriptscriptstyle{(1)}}  /2}^{\scriptscriptstyle{(1)}}(z)\cdot V_{\alpha^{\scriptscriptstyle{(2)}}+m/2b^{\scriptscriptstyle{(2)}}}^{\scriptscriptstyle{(2)}}(z),
\end{equation}
where by $V_{\alpha}^{\scriptscriptstyle{(\sigma)}}$  for $\sigma=1,2$ we denoted primary operator \eqref{primary} constructed for one of two subalgebras $\mathcal{H}\oplus\textsf{Vir}$:
\begin{equation*}
(\mathcal{H}\oplus\textsf{Vir})_{\sigma}\subset\mathcal{H}\oplus\mathcal{H}\oplus\mathcal{F}\oplus \textsf{NSR}.
\end{equation*}
We have checked equality \eqref{main-prop-equality} by explicit computations on lower levels. For further confirmations see appendix \ref{2Liouville}.

For practical purposes it is also useful to compute the ratio of the matrix elements (blow-up factors)
\begin{equation}\label{blowup-factors-def}
 l(\alpha,m|P',k',P,k)\overset{\text{def}}{=}
 \begin{cases}
    \frac{\langle k',P'|\mathbb{V}_{\alpha}^{(m)}|P,k\rangle}{\langle P'|\mathbb{V}_{\alpha}^{(0)}|P\rangle},\quad\text{if}\quad k+k'+m=2n,\\
     \frac{\langle k',P'|\mathbb{V}_{\alpha}^{(m)}|P,k\rangle}{\langle P'|\mathbb{V}_{\alpha}^{(\pm1)}|P\rangle},\quad\text{if}\quad k+k'+m=2n+1.
 \end{cases}
\end{equation}
\begin{Proposition}\label{blowup-Proposition}
The factors \eqref{blowup-factors-def} are given by
\begin{equation}
  l(\alpha,m|P',k',P,k)=
  \begin{cases}
  \prod_{i,j}
  s_{\textrm{even}}\left(\alpha+P'_{i}+P_{j},\frac{m+k'_{i}+k_{j}}{2}\right)\quad \qquad\;\;\,\text{if}\quad m+k+k'\quad\text{is even}\\
  \prod_{i,j}
  s_{\textrm{odd}}\left(\alpha+P'_{i}+P_{j},\mathrm{int}\,\Bigl(\frac{m+k'_{i}+k_{j}}{2}\Bigr)\right),
  \quad \text{if}\quad m+k+k'\quad\text{is odd}
  \end{cases}
\end{equation}
where $\vec{P}=(P,-P)$, $\vec{k}=(k,-k)$, $\vec{P}'=(P',-P')$, $\vec{k}'=(k',-k')$ and $\mathrm{int}(x)=\textrm{sgn}(x)\lfloor|x|\rfloor$ is the integer part of $x$ and for $n\geq0$
\begin{equation*}
\begin{aligned}
  &s_{\textrm{even}}(x,n)=2^{-\frac{n^{2}}{2}}\hspace*{-10pt}
  \prod_{\substack{i,j\geq1,\;i+j\leq2n\\i+j\equiv0\mod 2}}\hspace*{-10pt}(x+(i-1)b+(j-1)b^{-1}),\\
  &s_{\textrm{odd}}(x,n)=2^{-\frac{n(n+1)}{2}}\hspace*{-10pt}
  \prod_{\substack{i,j\geq1,\;i+j\leq2n+1\\i+j\equiv1\mod 2}}\hspace*{-10pt}(x+(i-1)b+(j-1)b^{-1}),
\end{aligned}
\end{equation*}
while for $n<0$ we have
\begin{equation*}
   s_{\textrm{even}}(x,n)=(-1)^{n}\,s_{\textrm{even}}(Q-x,-n),\quad
   s_{\textrm{odd}}(x,n)=s_{\textrm{odd}}(Q-x,-n).
\end{equation*}
\end{Proposition}
The proof of this proposition can be done by Coulomb integrals method and will be published elsewhere (see also appendix \ref{2Liouville}).
\section{Supersymmetric case: another compactification}\label{SAGT-collored}
The basis constructed in section \ref{SAGT} corresponds to the manifold of moduli of framed torsion free sheaves on $X_2$. As was mentioned in the Introduction there is another partial  compactification of the moduli space of instantons on  $\mathbb{C}^{2}/{\mathbb{Z}_{2}}$. This compactification will be explored in this section.
\subsection{Another compactification}
Recall that $\mathcal{M}(r,N)$ denotes the compactified moduli spaces of $U(r)$ instantons on $\mathbb{C}^{2}$  with the instanton number $N$. For any numbers $q_1,q_2,\dots q_r=0,1$ there is a natural action of $\mathbb{Z}_2$ on $\mathcal{M}(r,N)$:
\begin{align*}
B_{1} \mapsto -B_{1};\quad
B_{2}=-B_{2} ;\quad
I =  Iq;\quad J=qJ,
\end{align*}
where $q=\textrm{diag}((-1)^{q_{1}},\dots,(-1)^{q_{r}})$. Denote by $\mathcal{M}(r,N)^{\mathbb{Z}_2}$ the $\mathbb{Z}_2$ invariant part of $\mathcal{M}(r,N)$.

The manifold $\mathcal{M}(r,N)^{\mathbb{Z}_2}$ is smooth but not connected. In order to describe connected components consider the $N$--dimensional tautological vector bundle $\mathcal{V}$ on $\mathcal{M}(r,N)$. Its fiber at the point $p=(B_1,B_2,I,J)$ coincides with the vector space $V$ obtained from the vectors $I_1,\dots,I_r$ by action of an algebra generated by the  operators $B_1$ and $B_2$. If $p \in \mathcal{M}(r,N)^{\mathbb{Z}_2}$ then $\mathbb{Z}_2$ acts on the fiber of $\mathcal{V}$ at $p$. Then $V$ can be decomposes $V_+\oplus V_-$, where $V_+$ is the trivial representation and $V_-$ is the sign representations of $\mathbb{Z}_2$. Two points $p$, $q$ belong to the same component if the dimensions of $V_+$ at these points coincide. We denote connected components as $\mathcal{M}(r,d,N)$ where $d=N_+-N_-$, and  $N_+$, $N_-$ equal to the ranks of the bundles $\mathcal{V}_+$ and $\mathcal{V}_-$ respectively\footnote{ The connectedness of $\mathcal{M}(r,d,N)$ follows from its description in terms of Nakajima quiver varieties.}. It is evident that $d \equiv N\, (\mathrm{mod}\, 2)$.

Torus action on $\mathcal{M}(r,N)^{\mathbb{Z}_2}$ is given by formula \eqref{torus-action}. Points $p_{\scriptscriptstyle{\vec{W}}}$ fixed under the torus action are labeled by the $r$-tuples of Young diagrams $\vec{W}=(W_1,\dots,W_{r})$. It is convenient to color these diagrams as follows: the box $s\in W_k$ with coordinates $(i,j)$ is white if $i-j+q_k \equiv 0\,(\mathrm{mod}\,2)$ and black otherwise. The numbers $N_+$ and $N_-$ equal to the number of white and black boxes respectively.

The determinant of the vector field $v=(\epsilon_{1},\epsilon_{2},a)$ at the fixed point $p_{\scriptscriptstyle{\vec{W}}}$ equals to \cite{Fucito:2004ry,Fucito:2006kn}
\begin{equation}\label{eq det2 color}
  \det v\Bigl|_{p_{\scriptscriptstyle{\vec{W}}}}= Z_{\textsf{vec}}^{\diamond}(\vec{a},\vec{W}|\epsilon_{1},\epsilon_{2})^{-1}=
  \prod_{i,j=1}^{2}
 \prod_{s\in \scriptscriptstyle{W_{i}^{\diamond}}}
    E_{\scriptscriptstyle{W_{i}},\scriptscriptstyle{W_{j}}}(a_{i}-a_{j}|s)
   \bigl(\epsilon_{1}+\epsilon_{2}-E_{\scriptscriptstyle{W_{i}},\scriptscriptstyle{W_{j}}}(a_{i}-a_{j}|s)\bigr),
\end{equation}
where the superscript $\diamond$ means that the product goes over boxes $s\in W_{i}$ satisfying
\begin{equation*}
    \mathrm{a}_{\scriptscriptstyle{W_{i}}}(s)+\mathrm{l}_{\scriptscriptstyle{W_{j}}}(s)+1+q_i-q_j \equiv 0 \, (\mathrm{mod}\, 2).
\end{equation*}

In this subsection we consider the $r=2$ case. Following \cite{Belavin:2011pp} we choose components $\mathcal{M}(2,0,N)$ for $(q_1,q_2)=(0,0)$ and $\mathcal{M}(2,-1,N)$ for $(q_1,q_2)=(1,1)$\footnote{Such components satisfy the condition $q_1+q_2+2(N_+-N_-)=0$ which can be interpreted as the vanishing of the first Chern class \cite{Fucito:2004ry}.}. One can compute the Nekrasov partition function for the pure $U(r)$ gauge theory on $\mathbb{C}^2/\mathbb{Z}_2$ using these components:
\begin{equation}\label{Zpure-def-col}
   Z_{\textrm{pure}}^{\diamond}(\vec{a},\epsilon_{1},\epsilon_{2}|\Lambda)=
   \sum_{k=0}^{\infty}\sum_{\diamond} Z_{\textsf{vec}}^{\diamond}(\vec{a},\vec{W}|\epsilon_{1},\epsilon_{2})\,\Lambda^{2k},
\end{equation}
where the second sum goes over pairs of diagrams $\vec{W}$ with $|W|=k$, $N_+=N_-$ and with white corners or over pairs of diagrams with $|W|=k$, $N_+=N_--1$ and with black corners. As it was conjectured and checked in \cite{Belavin:2011pp} this function coincides with Whittaker limit of the four-point conformal block in $\mathcal{N}=1$ supersymmetric conformal field theory.

From the other side it was conjectured and checked in \cite{Bonelli:2011jx} that the function $Z_{\textrm{pure}}^{(2,X_{2})}(\vec{a},\epsilon_{1},\epsilon_{2}|q)$ defined by \eqref{BMT-formula} coincides with the same conformal block as well. Hence, these partition functions equal to each other
\begin{align}
Z_{\textrm{pure}}^{\diamond}(\vec{a},\epsilon_{1},\epsilon_{2}|q)= Z_{\textrm{pure}}^{(2,X_{2})}(\vec{a},\epsilon_{1},\epsilon_{2}|q). \label{eq Z=Z}
\end{align}
Summands on the left hand side are labeled by pairs of colored Young diagrams $W_1,W_2$. Summands on the right hand side are labeled by pair of pairs of Young diagrams $\vec{Y}^{\scriptscriptstyle{(\sigma)}}=(Y_{1}^{\scriptscriptstyle{(\sigma)}},Y_{2}^{\scriptscriptstyle{(\sigma)}})$, $\sigma=1,2$ and one integer number $k\in\mathbb{Z}$. There exists a bijection between these two types of combinatorial data (see for example \cite[Sec 1.1 Ex. 8]{Macdonald} or \cite{Fucito:2006kn}). However the sets of summands on the left hand side and on the right hand side of \eqref{eq Z=Z} are different (see appendix \ref{Bersh-app}). The identity \eqref{eq Z=Z} is nontrivial, we have the equality of sums of different rational functions.

The formula \eqref{eq Z=Z} follows from the fact that for $N\in \mathbb{Z}$ manifolds $\mathcal{M}(X_{2},2,N)$ and $\mathcal{M}(2,0,2N)$ are the compactifications of the same manifold (moduli space of instantons on $\mathbb{C}^2/\mathbb{Z}_2$). Hence the integrals of the equivariant forms should be equal. Similarly for $N \in \mathbb{Z}+\frac12$ integrals over $\mathcal{M}(X_{2},2,N)$ and $\mathcal{M}(2,-1,2N)$ should be equal (see also \cite{1166.14007} and \cite{Nagao_2007}).

Geometrical arguments from the Introduction suggest the existence of the basis labeled by pair of colored Young diagrams in the representation of the algebra $\mathcal{H}\oplus\mathcal{H}\oplus\mathcal{F}\oplus \textsf{NSR}$. In notation for this basis we use superscript $\diamond$: $|P\rangle^{\diamond}_{\vec{W}}$. Norm of the vector $|P\rangle^{\diamond}_{\vec{W}}$ should equal to the $Z_{\textsf{vec}}^{\diamond}(\vec{a},\vec{W}|\epsilon_{1},\epsilon_{2})^{-1}$. The basis $|P\rangle^{\diamond}_{\scriptscriptstyle{\vec{W}}}$ differs from the basis $|P,k\rangle_{\scriptscriptstyle{\vec{Y}^{1},\vec{Y}^{2}}}$ constructed in Section 3 since sets of summands in \eqref{eq Z=Z} are different.

Although we do not have an explicit construction of such basis, we suggest the formula for matrix element of the vertex operator $\mathbb{V}^{(0)}_{\alpha}$  \eqref{super-primary} in this basis
\begin{align}
\frac{ _{\vec{\scriptscriptstyle{W}}}^{\diamond}\langle P'|\mathbb{V}^{(0)}_{\alpha}|P\rangle_{ \vec{\scriptscriptstyle{Y}}}^{\diamond}}
    {^{\diamond}\langle P'|\mathbb{V}^{(0)}_{\alpha}|P\rangle^{\diamond}}=
    Z_{\textsf{bif}}^\diamond(\alpha;\vec{P}',\vec{W};\vec{P},\vec{Y}|b,b^{-1}), \label{eq-matr-elem-color}
\end{align}
where
\begin{align*}
   Z_{\textsf{bif}}^\diamond(m;\vec{a}',\vec{W};\vec{a},\vec{Y}|\epsilon_{1},\epsilon_{2})=\prod_{i,j=1}^{r}
    \prod_{\diamond}\left(\epsilon_{1}+\epsilon_{2}-E_{\scriptscriptstyle{Y_{i}},\scriptscriptstyle{W_{j}}}(a_{i}-a'_{j}|s)-m\right)
    \prod_{\diamond}\left(E_{\scriptscriptstyle{W_{j}},\scriptscriptstyle{Y_{i}}}(a'_{j}-a_{i}|t)-m\right)
\end{align*}
and the product goes over boxes $s\in Y_{i}$ and $t\in W_{j}$ satisfying
\begin{equation*}
    \mathrm{a}_{\scriptscriptstyle{Y_{i}}}(s)+\mathrm{l}_{\scriptscriptstyle{W_{j}}}(s)+1+q_{\scriptscriptstyle{Y_i}}-
    q_{\scriptscriptstyle{W_j}} \equiv 0 \, (\mathrm{mod}\, 2); \quad
    \mathrm{a}_{\scriptscriptstyle{W_j}}(t)+\mathrm{l}_{\scriptscriptstyle{Y_{i}}}(t)+1+
    q_{\scriptscriptstyle{W_j}}-q_{\scriptscriptstyle{Y_i}} \equiv 0 \, (\mathrm{mod}\, 2).
\end{equation*}
We have checked the formula \eqref{eq-matr-elem-color} computing the five-point conformal block
$$
\langle P'|\mathbb{V}^{(0)}_{\alpha}(q_1)\mathbb{V}^{(0)}_{\alpha}(q_1q_2)\mathbb{V}^{(0)}_{\alpha}(1)|P\rangle,
$$
using two different bases, i.e. comparing in lowest orders in $q_{1}$ and $q_{2}$ the results obtained  with the help of \eqref{main-prop-equality} and \eqref{eq-matr-elem-color}.

We note that \eqref{eq-matr-elem-color} can be considered as a system of equations for unknown basis vectors $|P\rangle_{ \vec{\scriptscriptstyle{Y}}}^{\diamond}$. Unfortunately, the solution of this system is not unique. This is closely related to the fact that the vertex operator $\mathbb{V}^{(0)}_{\alpha}$ does not depend on $\widehat{\mathfrak{sl}}(2)_{2}$ bosons $h_{n}$. Additional constraints could be explicit expressions for matrix elements of  operators different from $\mathbb{V}_{\alpha}^{(0)}$. It is unlikely that the matrix elements  of the operators $\mathbb{V}_{\alpha}^{(m)}$ introduced in section \ref{SAGT} have nice factorized form similar to \eqref{eq-matr-elem-color} for $m\neq0$. 

Note that if $\epsilon_1+\epsilon_2=0$ (in CFT notations $Q=0$) the equality \eqref{eq Z=Z} become trivial. Geometrically it is related to the fact that the manifolds $\mathcal{M}(X_{2},2,N)$ and $\mathcal{M}(2,0,2N)$ are $\mathbb{C}^*$--diffeomorphic, where $\mathbb{C}^*$ acts on $\mathbb{C}^2$ by formula  $(z_1,z_2) \mapsto (wz_1,w^{-1}z_2)$. However these manifolds are not diffeomorphic as $\left(\mathbb{C}^*\right)^2$--manifolds because the determinants at fixed points are different.


\subsection{The $r=1$ case}
In this subsection we discuss the phenomena of existence of different bases mentioned above. For simplicity we restrict ourself to the case $r=1$.

Denote by $\mathcal{M}(X_{2},1,N)$ the moduli space of framed torsion free sheaves on $X_{2}$ of rank $1$ with Chern classes $c_{1}=0$, $c_{2}=N$. Torus fixed points are labeled by pairs of Young diagrams $(Y^{\scriptscriptstyle{(1)}},Y^{\scriptscriptstyle{(2)}})$, $|Y^{\scriptscriptstyle{(1)}}|+|Y^{\scriptscriptstyle{(2)}}|=N$ and the determinant of the vector field $v=(\epsilon_{1},\epsilon_{2},a)$ at the fixed point $p_{\scriptscriptstyle Y^{\scriptscriptstyle{(1)}},\scriptscriptstyle Y^{\scriptscriptstyle{(2)}}}$ equals to (see \cite{2011CMaPh.304..395B}):
\begin{equation}\label{eq-Zvec2}
   \det v\Bigl|_{p_{\scriptscriptstyle Y^{\scriptscriptstyle{(1)}},\scriptscriptstyle Y^{\scriptscriptstyle{(2)}}}}=
   Z_{\textsf{vec}}(Y^{\scriptscriptstyle{(1)}},Y^{\scriptscriptstyle{(2)}}|\epsilon_{1},\epsilon_{2})^{-1} =
   Z_{\textsf{vec}}(Y^{\scriptscriptstyle{(1)}}|2\epsilon_{1},\epsilon_{2}-\epsilon_{1})^{-1}
   Z_{\textsf{vec}}(Y^{\scriptscriptstyle{(2)}}|\epsilon_{1}-\epsilon_{2},2\epsilon_{2})^{-1},
\end{equation}
where $Z_{\textsf{vec}}$ is given in \eqref{Zvec-def} and we omit  $\vec{a}$  since in $r=1$ case $\vec{a}$ doesn't appear in formulas. Denote by
\begin{equation*}
\mathcal{Z}_N=\sum_{|Y^{\scriptscriptstyle{(1)}}|+|Y^{\scriptscriptstyle{(2)}}|=N}
Z_{\textsf{vec}}(Y^{\scriptscriptstyle{(1)}},Y^{\scriptscriptstyle{(2)}}|\epsilon_{1},\epsilon_{2}).
\end{equation*}
the coefficient in Nekrasov partition function. The expression $\mathcal{Z}_{N}$ equals to the integral over moduli space $\mathcal{M}(X_{2},1,N)$. From the general scheme it follows that there should be a basis labeled by $(Y^{\scriptscriptstyle{(1)}},Y^{\scriptscriptstyle{(2)}})$ in representation of the algebra $ \mathcal{H} \oplus \mathcal{H}$ (see \eqref{scheme-2}). The algebraic construction of this basis is similar to one given in Section \ref{SAGT}.

From the colored partition side consider all components $\mathcal{M}(1,d,N)$ (with $q_1=0$). The torus fixed points $p_{\scriptscriptstyle W} \in \mathcal{M}(1,d,N)$ are labeled by colored Young diagrams $W$ with $d(W)=d$, $|W|=N$. The determinant of the vector field $v=(\epsilon_{1},\epsilon_{2},a)$ at the fixed point $p_{\scriptscriptstyle W}$ equals to \cite{Fucito:2004ry,Fucito:2006kn}
\begin{equation}\label{eq det1 color}
  \det v\Bigl|_{ p_{\scriptscriptstyle W}}= Z_{\textsf{vec}}^{\diamond}(a,\vec{W}|\epsilon_{1},\epsilon_{2})^{-1}=
 \prod_{s\in \scriptscriptstyle{W^{\diamond}}}
    E_{\scriptscriptstyle{W},\scriptscriptstyle{W}}(0|s)
   \bigl(\epsilon_{1}+\epsilon_{2}-E_{\scriptscriptstyle{W},\scriptscriptstyle{W}}(0|s)\bigr),
\end{equation}
where the product goes over boxes $s\in W$ satisfying  $\mathrm{a}_{\scriptscriptstyle{W}}(s)+\mathrm{l}_{\scriptscriptstyle{W}}(s)+1 \equiv 0 \, (\mathrm{mod}\, 2).$

\begin{figure}
\psfrag{v}{$\varnothing$}
\psfrag{d}{$\scriptstyle d=0$}
\psfrag{d1}{$\scriptstyle d=1$}
\psfrag{d2}{$\scriptstyle d=-1$}
\psfrag{d3}{$\scriptstyle d=-2$}
\psfrag{d4}{$\scriptstyle d=2$}
\psfrag{l1}{\small level $0$}
\psfrag{l2}{\small level $1$}
\psfrag{l3}{\small level $2$}
\psfrag{l4}{\small level $3$}
\psfrag{l5}{\small level $4$}
	\centering
	\includegraphics[width=.7\textwidth]{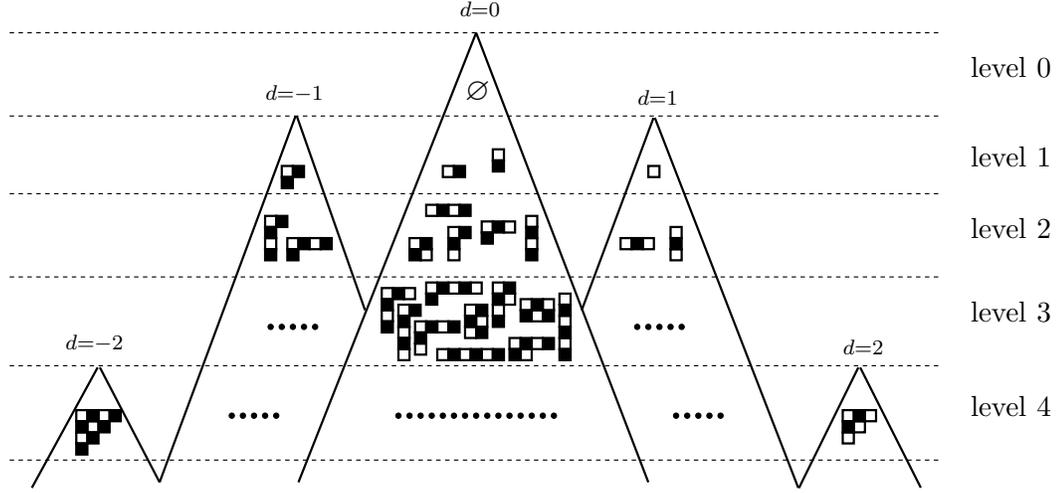}
        \caption{The colored partition basis in the representation of $\mathcal{H}\oplus \widehat{\mathfrak{sl}}(2)_1$. The interior of each angle  corresponds to the representation of $\mathcal{H}\oplus\mathcal{H}\subset\mathcal{H}\oplus\widehat{\mathfrak{sl}}(2)_{1}$ with given value of $h_{0}$. Each colored diagram represents a vector in this representation.}
	\label{CP-decompos-pic}
\end{figure}

Vectors $v_{\scriptscriptstyle W}$ corresponding to $p_{\scriptscriptstyle{W}}$ form a basis in representation of the  algebra $\mathcal{H}\oplus\widehat{\mathfrak{sl}}(2)_1$ (see \eqref{scheme-1}). Combinatorial gradings $d(W)$ and $|W|$ coincide with  $h_0$ grading and principal grading of representation of this algebra.
The structure of  representation of the algebra $\mathcal{H}\oplus\widehat{\mathfrak{sl}}(2)_1$ is shown on fig. \ref{CP-decompos-pic}.

Generators $e_{i}$ from $\widehat{\mathfrak{sl}}(2)_{1}$ shift $d$ by $+1$, generators $f_{i}$ by $-1$ and generators $h_{i}$ act in subspace with given $d$. Elements $h_{i}$ generate the Heisenberg algebra $\mathcal{H}\subset\widehat{\mathfrak{sl}}(2)_{1}$.

The vectors $v_{\scriptscriptstyle W}$ with given $d(W)=d$ form a basis in representation of the  algebra $ \mathcal{H} \oplus \mathcal{H}$. It is easy to see that the smallest diagram $W_0$ with $d(W_0)=d$ consist of $2d^2-d$ boxes and has a ``triangular'' form with edge length  $2|d|$ for $d \leq 0$ and $2d-1$ for $d>0$
\begin{equation}\label{2triangles}
\unitlength 2.3pt
\begin{picture}(130,20)(0,15)
\Thicklines
\path(0,0)(0,30)(30,30)(30,25)(25,25)(25,20)(20,20)(20,15)(15,15)(15,10)(10,10)(10,5)(5,5)(5,0)(0,0)
 \put(-7,19){\vector(0,1){11}}
 \put(-10,14){\mbox{\small{$2|d|$}}}
 \put(-7,11){\vector(0,-1){11}}
 \put(0,-11){\mbox{for $d<0$}}
\blacken\path(0,0)(0,5)(5,5)(5,0)(0,0)
\blacken\path(5,5)(5,10)(10,10)(10,5)(5,5)
\blacken\path(10,10)(10,15)(15,15)(15,10)(10,10)
\blacken\path(15,15)(15,20)(20,20)(20,15)(15,15)
\blacken\path(20,20)(20,25)(25,25)(25,20)(20,20)
\blacken\path(25,25)(25,30)(30,30)(30,25)(25,25)
\blacken\path(0,10)(0,15)(5,15)(5,10)(0,10)
\blacken\path(5,15)(5,20)(10,20)(10,15)(5,15)
\blacken\path(10,20)(10,25)(15,25)(15,20)(10,20)
\blacken\path(15,25)(15,30)(20,30)(20,25)(15,25)
\blacken\path(0,20)(0,25)(5,25)(5,20)(0,20)
\blacken\path(5,25)(5,30)(10,30)(10,25)(5,25)
\path(100,5)(100,30)(125,30)(125,25)(120,25)(120,20)(115,20)(115,15)(110,15)(110,10)(105,10)(105,5)(100,5)
\put(93,21){\vector(0,1){9}}
 \put(86,16){\mbox{\small{$2d-1$}}}
 \put(93,13){\vector(0,-1){9}}
 \put(100,-11){\mbox{for $d>0$}}
\blacken\path(100,10)(100,15)(105,15)(105,10)(100,10)
\blacken\path(105,15)(105,20)(110,20)(110,15)(110,15)
\blacken\path(110,20)(110,25)(115,25)(115,20)(115,20)
\blacken\path(115,25)(115,30)(120,30)(120,25)(120,25)
\blacken\path(100,20)(100,25)(105,25)(105,20)(100,20)
\blacken\path(105,25)(105,30)(110,30)(110,25)(110,25)
\end{picture}
\vspace*{2.5cm}
\end{equation}

Denote  by
\begin{equation*}
Z_{d,N}=\sum_{W,\, d(W)=d,\, |W|=N} Z_{\textsf{vec}}^{\diamond}(W|\epsilon_{1},\epsilon_{2})
\end{equation*}
the coefficient in the Nekrasov partition function. The expression $Z_{d,N}$ equals to the integral over moduli space $\mathcal{M}(1,d,N)$.
\begin{Proposition} For any integer $d$
\begin{align}Z_{d,2d^2-d+2N} \label{eq Z=Z=Z}=Z_{0,2N}=\mathcal{Z}_N\end{align}
\end{Proposition}
This proposition follows from the fact that the manifolds $\mathcal{M}(1,d,2d^2-d+2N)$ and  $\mathcal{M}(X_{2},1,N)$ are birationally isomorphic to the Hilbert scheme of $N$ point on $\mathbb{C}^2/\mathbb{Z}_2$.

The equality \eqref{eq Z=Z=Z} is an equality of sums. The number of summands from the left hand side and right hand side is the same (this follows from the bijection mentioned above). We will write $\sum \equiv \sum$ if sums are equal and moreover the sets of summands on both sides are the same. Correspondingly we will write  $\sum \not\equiv \sum$ if the sums are equal but the sets of summands are different. Direct calculations shows:
$$Z_{0,0}\equiv Z_{1,1} \equiv Z_{-1,3} \equiv Z_{2,6} \equiv Z_{-2,10} \equiv \mathcal{Z}_0.$$
$$Z_{0,2}\equiv Z_{1,3} \equiv Z_{-1,5} \equiv Z_{2,8} \equiv Z_{-2,12} \equiv \mathcal{Z}_1.$$
$$Z_{0,4}\equiv Z_{1,5} \equiv Z_{-1,7} \equiv Z_{2,10} \equiv Z_{-2,14} \equiv \mathcal{Z}_2.$$
$$Z_{0,6}\not\equiv  Z_{1,7},\quad Z_{1,7}\equiv Z_{-1,9} \equiv Z_{2,12} \equiv Z_{-2,16} \equiv \mathcal{Z}_3.$$
$$Z_{0,8}\not\equiv  Z_{1,9},\quad Z_{0,8}\not\equiv  Z_{-1,11},\quad Z_{1,9}\not\equiv  Z_{-1,11},\quad Z_{-1,11}\equiv Z_{2,14} \equiv Z_{-2,18} \equiv \mathcal{Z}_4.$$
\begin{gather*} Z_{0,10}\not\equiv  Z_{1,11},\quad Z_{0,10}\not\equiv  Z_{-1,13},\quad Z_{1,11}\not\equiv  Z_{-1,13},\\ Z_{0,10}\not\equiv  Z_{2,16},\quad Z_{1,11}\not\equiv  Z_{2,16},\quad Z_{-1,12}\not\equiv  Z_{2,16}, \quad Z_{2,16}\equiv Z_{-2,20} \equiv Z_{3,25} \equiv \mathcal{Z}_{5}.
\end{gather*}
These results suggest the following proposition\footnote{The same phenomena was independently noticed by R. Poghossian \cite{Pog-cite}.}
\begin{itemize}
\item For any $d_1,d_2$ there exists $N$ such that $Z_{d_1,2d_1^2-d_1+2N}\not\equiv Z_{d_2,2d_2^2-d_2+2N}$

\item For any $N$ there exist $D$ such that $Z_{d,2d^2-d+2N}\equiv\mathcal{Z}_N$ for any $d,\, |d|\geq D$.
\end{itemize}
In terms of basis this proposition means that there exists an infinite number of different bases in representation of the algebra $\mathcal{H} \oplus \mathcal{H}$. These bases are numbered by integer number $d$. Basic vectors in $d$-th basis are labeled by Young diagrams $W$ with $d(W)=d$. The basis labeled by pairs of Young diagrams $Y^{\scriptscriptstyle{(1)}},Y^{\scriptscriptstyle{(2)}}$ appears in the limit $d \rightarrow \infty$.

We prove the second assertion:
\begin{Proposition} If $|d| \geq N$, then
\begin{align}Z_{d,2d^2-d+2N}\equiv\mathcal{Z}_N\label{eq Z=Z limit}.
\end{align}
\end{Proposition}
The proof is based on the explicit bijection: for any pair of Young diagram $Y^{\scriptscriptstyle{(1)}},Y^{\scriptscriptstyle{(2)}}$ with $|Y^{\scriptscriptstyle{(1)}}|+|Y^{\scriptscriptstyle{(2)}}|=N$ we construct colored Young diagram
$W$ with $|W|=2d^2-d+2N$, $d(W)=d$ such that
\begin{equation}\label{eq Z=Z biject}
 Z_{\textsf{vec}}^{\diamond}(W|\epsilon_{1},\epsilon_{2})=
 Z_{\textsf{vec}}(Y^{\scriptscriptstyle{(1)}},Y^{\scriptscriptstyle{(2)}}|\epsilon_{1},\epsilon_{2}).
\end{equation}

\begin{figure}
\psfrag{y1}{$Y^{\scriptscriptstyle{(1)}}$}
\psfrag{y2}{$Y^{\scriptscriptstyle{(2)}}$}
\psfrag{yy1}{$\widetilde{Y}^{\scriptscriptstyle{(1)}}$}
\psfrag{yy2}{$\widetilde{Y}^{\scriptscriptstyle{(2)}}$}
\psfrag{w}{$W_{0}$}
\psfrag{ww}{$W$}
	\centering
	\includegraphics[width=.86\textwidth]{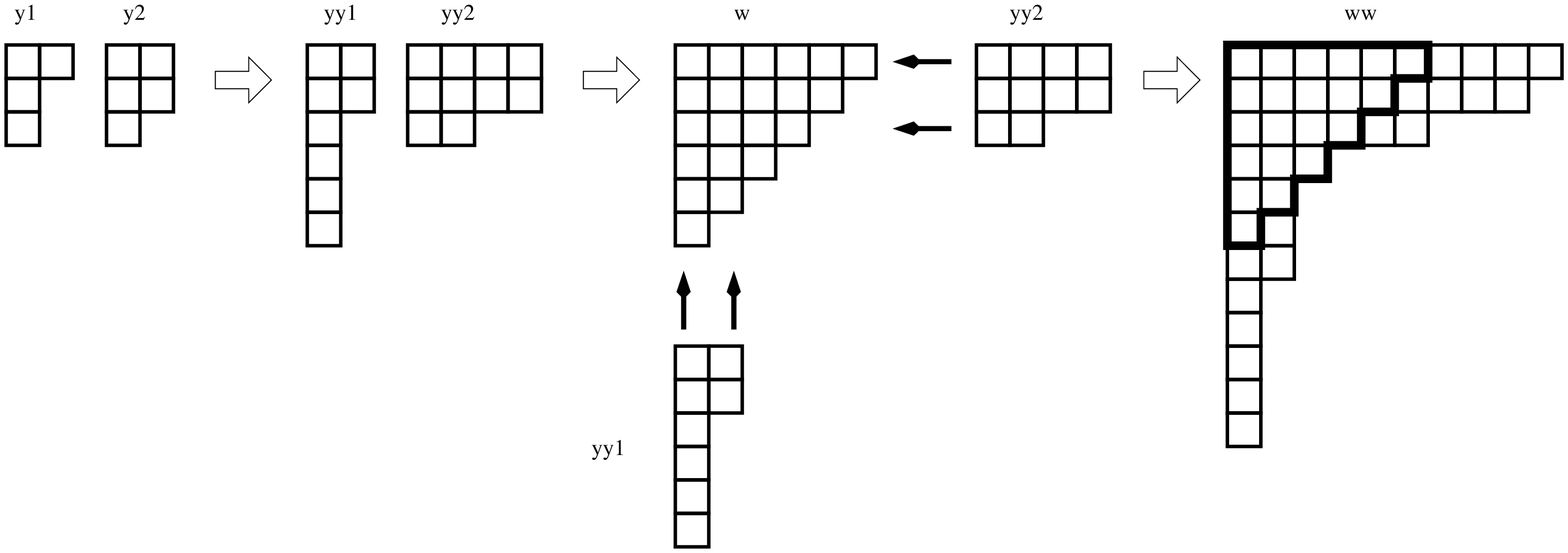}
        \caption{Bijection between the pair $(Y^{\scriptscriptstyle{(1)}},Y^{\scriptscriptstyle{(2)}})$ and $W$.}
	\label{bijection}
\end{figure}

Bijection goes as follows. Denote by $W_0$ the minimal Young diagram with $d(W_0)=d$. Then $|W_0|=2d^2-d$ and $W_0$ has ``triangular'' form \eqref{2triangles}. By $\widetilde{Y}^{\scriptscriptstyle{(1)}}$ denote diagram obtained from $Y^{\scriptscriptstyle{(1)}}$ by doubling all columns. Similarly, by $\widetilde{Y}^{\scriptscriptstyle{(2)}}$ denote diagram obtained from $Y^{\scriptscriptstyle{(2)}}$ by doubling all rows. Then $W$ is obtained by adding diagrams $\widetilde{Y}^{\scriptscriptstyle{(1)}}$ and $\widetilde{Y}^{\scriptscriptstyle{(2)}}$ to the bottom and to the right of the diagram $W_0$ respectively (see fig. \ref{bijection}).

The added diagrams $\widetilde{Y}^{\scriptscriptstyle{(1)}}$ and $\widetilde{Y}^{\scriptscriptstyle{(2)}}$ do not interact since $|d| \geq N$. Then, the identity \eqref{eq Z=Z biject} follows from easy combinatorics. $\square$

In this subsection we considered the $r=1$ case only. For general $r$ situation is quite similar, there should be a sequence of bases labelled by integer number $d$. The basis corresponding to $\mathcal{M}_2(r,N)$ appears in the limit $d \rightarrow \infty$.
\section{Concluding remarks}\label{Concl}
\begin{enumerate}
\item It would be interesting to give an explicit construction of the basis
labeled by colored partitions. As we see in section \ref{SAGT-collored} this basis is not determined by formula for the matrix element \eqref{eq-matr-elem-color}.
\item It would be interesting to generalize results of sections \ref{SAGT} and \ref{SAGT-collored} for the general case $p>2$. Note that on the instanton moduli side this case is very similar to the $p=2$ case.
\end{enumerate}
\section*{Acknowledgments}
We thank Giulio Bonelli, Ivan Cherednik, Vladimir Fateev, Anatol Kirillov, Alexander Kuznetsov, Rubik Poghossian and Alessandro Tanzini for discussions.  We also thank Slava Pugai and Rubik Poghossian for critical reading of the manuscript and very useful comments.
Some of authors (A.B., M.B. and A.L.) are grateful for organizers of the workshop ``Low dimensional physics and gauge principles'' held in Nor Amberd, Armenia in September 2011 for hospitality and stimulating scientific atmosphere. A.B. thanks Boris Dubrovin for kind hospitality during his visit to SISSA and for interesting discussions.

This research was held within the framework of the Federal programs ``Scientific and Scientific-Pedagogical Personnel of Innovational Russia'' on 2009-2013 (state contracts No. P1339 and No. 02.740.11.5165) and was supported by RFBR  grants 12-01-00836 and 12-02-01092 and by Russian Ministry of Science and Technology under the Scientific Schools grant 6501.2010.2.  The research of A.L. and G.T. was also supported in part by the National Science Foundation under Grant No. NSF PHY05-51164 and by Dynasty foundation.

\Appendix
\section{More on two Virasoro algebras in $\mathcal{F}\oplus\mathsf{NSR}$}\label{2Liouville}
In section \ref{SAGT} we observed ``strange'' relation \eqref{strange-relation} which is equivalent to\footnote{For $m=0$ this relation was noticed in \cite{Crnkovic:1989ug}.}
\begin{equation}\label{des-fields}
\begin{aligned}
 &\Phi_{\alpha}^{\scriptscriptstyle{\textsf{NS}}}(z)\simeq
 V_{\alpha^{\scriptscriptstyle{(1)}}}^{\scriptscriptstyle{\textsf{Vir}_{1}}}(z)\cdot V_{\alpha^{\scriptscriptstyle{(2)}}}^{\scriptscriptstyle{\textsf{Vir}_{2}}}(z),\\
 &\alpha f(z)\Phi_{\alpha}^{\scriptscriptstyle{\textsf{NS}}}(z)+
   \Psi_{\alpha}^{\scriptscriptstyle{\textsf{NS}}}(z)\simeq
   V_{\alpha^{\scriptscriptstyle{(1)}}+b^{\scriptscriptstyle{(1)}}  /2}^{\scriptscriptstyle{\textsf{Vir}_{1}}}(z)\cdot V_{\alpha^{\scriptscriptstyle{(2)}}+1/2b^{\scriptscriptstyle{(2)}}  }^{\scriptscriptstyle{\textsf{Vir}_{2}}}(z),\\
  &(Q-\alpha) f(z)\Phi_{\alpha}^{\scriptscriptstyle{\textsf{NS}}}(z)+
   \Psi_{\alpha}^{\scriptscriptstyle{\textsf{NS}}}(z)\simeq
   V_{\alpha^{\scriptscriptstyle{(1)}}-b^{\scriptscriptstyle{(1)}}  /2}^{\scriptscriptstyle{\textsf{Vir}_{1}}}(z)\cdot V_{\alpha^{\scriptscriptstyle{(2)}}-1/2b^{\scriptscriptstyle{(2)}}}^{\scriptscriptstyle{\textsf{Vir}_{2}}}(z),\\
   &\dots\dots\dots\dots\dots\dots\dots\dots\dots\dots\dots\dots\dots\dots\dots\dots
\end{aligned}
\end{equation}
i.e. for any $m\in\mathbb{Z}$ the product
\begin{equation*}
 V_{\alpha^{\scriptscriptstyle{(1)}}+mb^{\scriptscriptstyle{(1)}}  /2}^{\scriptscriptstyle{\textsf{Vir}_{1}}}(z)\cdot V_{\alpha^{\scriptscriptstyle{(2)}}+m/2b^{\scriptscriptstyle{(2)}}  }^{\scriptscriptstyle{\textsf{Vir}_{2}}}(z)
\end{equation*}
of two primary fields in two CFT's $\textsf{Vir}_{1}$ and $\textsf{Vir}_{2}$ with parameters satisfying \eqref{bb-relat-1}
is equal up to normalization to the descendant field on level $m^{2}/2$ of the field $\Phi_{\alpha}^{\scriptscriptstyle{\textsf{NS}}}(z)$ in $\mathcal{F}\oplus\textsf{NSR}$ theory. In operator language this descendant  field corresponds to the highest weight vector \eqref{high-weight-def}.
First check which we can perform is to compare conformal dimensions. One can easily find that
\begin{equation*}
   \Delta(\alpha^{\scriptscriptstyle{(1)}}+mb^{\scriptscriptstyle{(1)}}  /2,b^{\scriptscriptstyle{(1)}}  )+
   \Delta(\alpha^{\scriptscriptstyle{(2)}}+m/2b^{\scriptscriptstyle{(2)}}  ,b^{\scriptscriptstyle{(2)}}  )=\Delta_{\textsf{\tiny{NS}}}(\alpha,b)+\frac{m^{2}}{2},
\end{equation*}
where $\Delta(\alpha,b)$ and $\Delta_{\textsf{\tiny{NS}}}(\alpha,b)$ are the conformal dimensions parameterized in Virasoro (Liouville) \eqref{Delta-Vir} and \eqref{Delta-NS} NSR (Super Liouville) manners.

Another more concrete check would be to compare three-point correlation functions. We consider the relation (other relations \eqref{des-fields} can be treated similarly)
\begin{equation}\label{FF-VV}
 \Phi_{\alpha}^{\scriptscriptstyle{\textsf{NS}}}(z)\simeq
 V_{\alpha^{\scriptscriptstyle{(1)}}}^{\scriptscriptstyle{\textsf{Vir}_{1}}}(z)\cdot V_{\alpha^{\scriptscriptstyle{(2)}}}^{\scriptscriptstyle{\textsf{Vir}_{2}}}(z).
 \end{equation}
Right hand side of \eqref{FF-VV} is given by the products of two primary operators in two CFT's $\textsf{Vir}_{1}$ and $\textsf{Vir}_{2}$ with central charges $c^{\scriptscriptstyle{(1)}}$ and $c^{\scriptscriptstyle{(2)}}$ parameterized as
\begin{equation*}
  c^{\scriptscriptstyle{(\sigma)}}=1+6\left(b^{\scriptscriptstyle{(\sigma)}}+\frac{1}{b^{\scriptscriptstyle{(\sigma)}}}\right)^{2},
\end{equation*}
where $b^{\scriptscriptstyle{(\sigma)}}$ are given by
\begin{equation*}
b^{\scriptscriptstyle{(1)}}  =\frac{2b}{\sqrt{2-2b^{2}}},\quad
   (b^{\scriptscriptstyle{(2)}})^{-1}=\frac{2b^{-1}}{\sqrt{2-2b^{-2}}}.
\end{equation*}
Let us consider the region $b<1$. In this case $b^{\scriptscriptstyle{(1)}}  $ is real while $b^{\scriptscriptstyle{(2)}}  $ is imaginary. For general values of all the parameters we can treat theories $\textsf{Vir}_{1}$ and $\textsf{Vir}_{2}$ as the Liouville field theory \cite{Zamolodchikov:1995aa} with coupling constant $b^{\scriptscriptstyle{(1)}}$ and generalized minimal model \cite{Zamolodchikov:2005fy} (time-like Liouville field theory) with coupling constant $\hat{b}^{\scriptscriptstyle{(2)}}$ (we have fixed the brunch cut as $b^{\scriptscriptstyle{(2)}}=-i\hat{b}^{\scriptscriptstyle{(2)}}$). The three-point functions in both theories
\begin{equation}
   \begin{aligned}
      &C(\alpha_{1}^{\scriptscriptstyle{(1)}},\alpha_{2}^{\scriptscriptstyle{(1)}},\alpha_{3}^{\scriptscriptstyle{(1)}}|b^{\scriptscriptstyle{(1)}})
      \overset{\text{def}}{=}
      \langle V_{\alpha_{1}^{\scriptscriptstyle{(1)}}}(0)V_{\alpha_{2}^{\scriptscriptstyle{(1)}}}(1)V_{\alpha_{3}^{\scriptscriptstyle{(1)}}}(\infty)
      \rangle_{b^{\scriptscriptstyle{(1)}}},\\
      &\hat{C}(\hat{\alpha}_{1}^{\scriptscriptstyle{(2)}},\hat{\alpha}_{2}^{\scriptscriptstyle{(2)}},\hat{\alpha}_{3}^{\scriptscriptstyle{(2)}}
      |\hat{b}^{\scriptscriptstyle{(2)}})\overset{\text{def}}{=}
      \langle V_{\hat{\alpha}_{1}^{\scriptscriptstyle{(2)}}}(0)V_{\hat{\alpha}_{2}^{\scriptscriptstyle{(2)}}}(1)V_{\hat{\alpha}_{3}^{\scriptscriptstyle{(2)}}}(\infty)\rangle_{\hat{b}^{\scriptscriptstyle{(2)}}},
   \end{aligned}
\end{equation}
where
\begin{equation}\label{alpha-sigma}
 \begin{aligned}
    &b^{\scriptscriptstyle{(1)}}  =\frac{2b}{\sqrt{2-2b^{2}}},&\quad &\alpha^{\scriptscriptstyle{(1)}}=\frac{\alpha}{\sqrt{2-2b^{2}}},\\
    &(\hat{b}^{\scriptscriptstyle{(2)}})^{-1}=\frac{2}{\sqrt{2-2b^{2}}},&\quad&\hat{\alpha}^{\scriptscriptstyle{(2)}}=\frac{b\alpha}{\sqrt{2-2b^{2}}}.
 \end{aligned}
\end{equation}
These three-point functions are known in explicit form \cite{Zamolodchikov:1995aa}
\begin{subequations}\label{C}
\begin{multline}\label{CL}
 C(\alpha_{1},\alpha_{2},\alpha_{3}|b)=
 \frac{\Upsilon_{b}(b)\Upsilon_{b}(2\alpha_{1})\Upsilon_{b}(2\alpha_{2})\Upsilon_{b}(2\alpha_{3})}
 {\Upsilon_{b}(\alpha_{1}+\alpha_{2}+\alpha_{3}-Q)\Upsilon_{b}(\alpha_{1}+\alpha_{2}-\alpha_{3})
 \Upsilon_{b}(\alpha_{1}+\alpha_{3}-\alpha_{2})
 \Upsilon_{b}(\alpha_{2}+\alpha_{3}-\alpha_{1})}
\end{multline}
and \cite{Zamolodchikov:2005fy}
\begin{multline}\label{CM}
 \hat{C}(\alpha_{1},\alpha_{2},\alpha_{3}|b)=\\=
 \frac{\Upsilon_{b}(b)\Upsilon_{b}(\alpha_{1}+\alpha_{2}+\alpha_{3}-b^{-1}+2b)\Upsilon_{b}(\alpha_{1}+\alpha_{2}-\alpha_{3}+b)
 \Upsilon_{b}(\alpha_{1}+\alpha_{3}-\alpha_{2}+b)
 \Upsilon_{b}(\alpha_{2}+\alpha_{3}-\alpha_{1}+b)}
 {\Upsilon_{b}(2\alpha_{1}+b)\Upsilon_{b}(2\alpha_{2}+b)\Upsilon_{b}(2\alpha_{3}+b)},
\end{multline}
\end{subequations}
where $\Upsilon_{b}(x)$ is the entire selfdual function (with respect to transformation $b\rightarrow1/b$), which was defined in \cite{Zamolodchikov:1995aa} by the integral representation
\begin{equation}
    \log\Upsilon_{b}(x)=\int_{0}^{\infty}\frac{dt}{t}
    \left[\left(\frac{b+b^{-1}}{2}-x\right)^2e^{-t}-\frac
    {\sinh^2\left(\frac{b+b^{-1}}{2}-x\right)\frac{t}{2}}
    {\sinh\frac{bt}{2}\sinh\frac{t}{2b}}
    \right].
\end{equation}
Equations \eqref{C} are written up to some factors which can be eliminated by changing of normalization of the primary operators which is always in our hands. For the fields in the left hand side of \eqref{FF-VV} we can define the three-point function
\begin{equation}
   C_{\scriptscriptstyle{\textsf{NS}}}(\alpha_{1},\alpha_{2},\alpha_{3})\overset{\text{def}}{=}
   \langle \Phi_{\alpha_{1}}^{\scriptscriptstyle{\textsf{NSR}}}(0)
   \Phi_{\alpha_{2}}^{\scriptscriptstyle{\textsf{NSR}}}(1)\Phi_{\alpha_{3}}^{\scriptscriptstyle{\textsf{NSR}}}(\infty)
      \rangle_{b},
\end{equation}
where the average is understood as an average in the Super-Liouville field theory with coupling constant $b$. Following \cite{Rashkov:1996jx,Poghosian:1996dw} it has the following explicit form (again up to normalization of primary fields)
\begin{equation}
   C_{\scriptscriptstyle{\textsf{NS}}}(\alpha_{1},\alpha_{2},\alpha_{3})=
 \frac{\Upsilon^{\scriptscriptstyle{\textsf{NS}}}_{b}(2\alpha_{1})\Upsilon^{\scriptscriptstyle{\textsf{NS}}}_{b}(2\alpha_{2})
 \Upsilon^{\scriptscriptstyle{\textsf{NS}}}_{b}(2\alpha_{3})}
 {\Upsilon^{\scriptscriptstyle{\textsf{NS}}}_{b}(\alpha_{1}+\alpha_{2}+\alpha_{3}-Q)
 \Upsilon^{\scriptscriptstyle{\textsf{NS}}}_{b}(\alpha_{1}+\alpha_{2}-\alpha_{3})
 \Upsilon^{\scriptscriptstyle{\textsf{NS}}}_{b}(\alpha_{1}+\alpha_{3}-\alpha_{2})
 \Upsilon^{\scriptscriptstyle{\textsf{NS}}}_{b}(\alpha_{2}+\alpha_{3}-\alpha_{1})},
\end{equation}
where
\begin{equation*}
 \Upsilon^{\scriptscriptstyle{\textsf{NS}}}_{b}(x)\overset{\text{def}}{=}
 \Upsilon_{b}\left(\frac{x}{2}\right)\Upsilon_{b}\left(\frac{x+Q}{2}\right).
\end{equation*}
Using the relation\footnote{We note that this relation is very similar to the relation used in ref. \cite{Bershtein:2010wz}, where the connection between the parafermionic Liouville theory and the three-exponential model \cite{Fateev:1996ea} was studied.}
\begin{equation}
   \frac{\Upsilon_{b^{\scriptscriptstyle{(1)}}}(\alpha^{\scriptscriptstyle{(1)}})}
   {\Upsilon_{\hat{b}^{\scriptscriptstyle{(2)}}}(\hat{\alpha}^{\scriptscriptstyle{(2)}}+\hat{b}^{\scriptscriptstyle{(2)}})}=
   \frac{\Upsilon_{b^{\scriptscriptstyle{(1)}}}(b^{\scriptscriptstyle{(1)}})}
   {\Upsilon_{\hat{b}^{\scriptscriptstyle{(2)}}}(\hat{b}^{\scriptscriptstyle{(2)}})\Upsilon_{b}(b)}\,b^{\frac{b^{2}\alpha(Q-\alpha)}{2-2b^{2}}}
   \left(\frac{1-b^{2}}{2}\right)^{\frac{\alpha(Q-\alpha)}{4}-\frac{1}{2}}
   \Upsilon^{\scriptscriptstyle{\textsf{NS}}}_{b}(\alpha),
\end{equation}
one can check that
\begin{equation}\label{CC=C}
  C(\alpha_{1}^{\scriptscriptstyle{(1)}},\alpha_{2}^{\scriptscriptstyle{(1)}},\alpha_{3}^{\scriptscriptstyle{(1)}}|b^{\scriptscriptstyle{(1)}})
  \hat{C}(\hat{\alpha}_{1}^{\scriptscriptstyle{(2)}},\hat{\alpha}_{2}^{\scriptscriptstyle{(2)}},\hat{\alpha}_{3}^{\scriptscriptstyle{(2)}}
      |\hat{b}^{\scriptscriptstyle{(2)}})\simeq
      C_{\scriptscriptstyle{\textsf{NS}}}(\alpha_{1},\alpha_{2},\alpha_{3}).
\end{equation}
We note that choosing appropriate normalization of the fields one can always set the coefficient of proportionality in \eqref{CC=C} to be equal to $1$.

The ratio of the matrix elements \eqref{blowup-factors-def} can also be interpreted within this framework. Namely, let us assume that $m+k+k'$ is an even number, then
\begin{multline}\label{l=CC}
  l(\alpha,m|P',k',P,k)^{2}\simeq
  \frac{C(\alpha_{1}^{\scriptscriptstyle{(1)}}+kb^{\scriptscriptstyle{(1)}}/2,\alpha_{2}^{\scriptscriptstyle{(1)}}+k'b^{\scriptscriptstyle{(1)}}/2,
  \alpha^{\scriptscriptstyle{(1)}}+mb^{\scriptscriptstyle{(1)}}/2|b^{\scriptscriptstyle{(1)}})}
  {C(\alpha_{1}^{\scriptscriptstyle{(1)}},\alpha_{2}^{\scriptscriptstyle{(1)}},\alpha^{\scriptscriptstyle{(1)}}|b^{\scriptscriptstyle{(1)}})}
  \times\\\times
   \frac{\hat{C}(\hat{\alpha}_{1}^{\scriptscriptstyle{(2)}}+k/2\hat{b}^{\scriptscriptstyle{(2)}},\hat{\alpha}_{2}^{\scriptscriptstyle{(2)}}+k'/2\hat{b}^{\scriptscriptstyle{(2)}},
  \hat{\alpha}^{\scriptscriptstyle{(2)}}+m/2\hat{b}^{\scriptscriptstyle{(2)}}
      |\hat{b}^{\scriptscriptstyle{(2)}})}
  {\hat{C}(\hat{\alpha}_{1}^{\scriptscriptstyle{(2)}},\hat{\alpha}_{2}^{\scriptscriptstyle{(2)}},\hat{\alpha}^{\scriptscriptstyle{(2)}}
      |\hat{b}^{\scriptscriptstyle{(2)}})},
\end{multline}
where
\begin{equation*}
   \alpha_{1}=\frac{Q}{2}+P,\qquad
   \alpha_{2}=\frac{Q}{2}+P',
\end{equation*}
and the sets $(\alpha_{1}^{\scriptscriptstyle{(\sigma)}},\alpha_{2}^{\scriptscriptstyle{(\sigma)}},\alpha^{\scriptscriptstyle{(\sigma)}})$ and $(\hat{\alpha}_{1}^{\scriptscriptstyle{(\sigma)}},\hat{\alpha}_{2}^{\scriptscriptstyle{(\sigma)}},\hat{\alpha}^{\scriptscriptstyle{(\sigma)}})$ are related to $(\alpha_{1},\alpha_{2},\alpha)$ as in \eqref{alpha-sigma}. Equation \eqref{l=CC} can be checked (again up to normalization of the fields) using the relation
\begin{multline}
   \frac{\Upsilon_{b^{\scriptscriptstyle{(1)}}}(\alpha^{\scriptscriptstyle{(1)}})}
   {\Upsilon_{b^{\scriptscriptstyle{(1)}}}(\alpha^{\scriptscriptstyle{(1)}}+nb^{\scriptscriptstyle{(1)}})}
   \frac{\Upsilon_{\hat{b}^{\scriptscriptstyle{(2)}}}(\hat{\alpha}^{\scriptscriptstyle{(2)}}+\hat{b}^{\scriptscriptstyle{(2)}}+n/\hat{b}^{\scriptscriptstyle{(2)}})}
   {\Upsilon_{\hat{b}^{\scriptscriptstyle{(2)}}}(\hat{\alpha}^{\scriptscriptstyle{(2)}}+\hat{b}^{\scriptscriptstyle{(2)}})}=
   \frac{(-1)^{n}}{(2-2b^{2})^{n^{2}}}\,
   b^{\frac{2bn}{(1-b^{2})}(x+nb^{-1}-Q/2)}
   \times\\\times
   \hspace*{-10pt}
  \prod_{\substack{i,j\geq1,\;i+j\leq2n\\i+j\equiv0\mod 2}}\hspace*{-10pt}(\alpha+(i-1)b+(j-1)b^{-1})^{2}.
\end{multline}
The case when $m+k+k'$ is an odd number can be treated similarly.
\section{Highest weight vectors}\label{Highest-weight}
In this appendix we give explicit expressions for the highest weight vectors $|P,k\rangle$ defined by \eqref{high-weight-def} with
\begin{equation}
  \Delta^{\scriptscriptstyle{(1)}}(P,k)=\frac{(Q^{\scriptscriptstyle{(1)}})^{2}}{4}-\left(P^{\scriptscriptstyle{(1)}}
  +\frac{kb^{\scriptscriptstyle{(1)}}}{2}\right)^{2},\quad
  \Delta^{\scriptscriptstyle{(2)}}(P,k)=\frac{(Q^{\scriptscriptstyle{(2)}})^{2}}{4}-\left(P^{\scriptscriptstyle{(2)}}
  +\frac{k}{2b^{\scriptscriptstyle{(2)}}}\right)^{2}.
\end{equation}
The state $|P,k\rangle$ belongs to the level $k^{2}/2$ of the highest weight representation $|P\rangle$ of the algebra $\mathcal{F}\oplus\textsf{NSR}$. For each value of $k^{2}/2$ there are exactly two states $|P,k\rangle$ and $|P,-k\rangle$ orthogonal to each other. For example, on the level $1/2$ we have
\begin{equation}
    \begin{aligned}
        &|P,1\rangle=\left(G_{-\frac{1}{2}}+(Q/2+P)f_{-\frac{1}{2}}\right)|P\rangle_{\scriptscriptstyle{\textsf{NS}}},\\
        &|P,-1\rangle=\left(G_{-\frac{1}{2}}+(Q/2-P)f_{-\frac{1}{2}}\right)|P\rangle_{\scriptscriptstyle{\textsf{NS}}}
    \end{aligned}
\end{equation}
and on the level $2$
\begin{subequations}
\begin{multline}
        |P,2\rangle=\bigl(G_{-\frac{1}{2}}^{4}+(Q/2+P)^{2}G_{-\frac{1}{2}}G_{-\frac{3}{2}}-
        (Q/2+P+b)(Q/2+P+b^{-1})G_{-\frac{3}{2}}G_{-\frac{1}{2}}
        -2(Q+P)G_{-\frac{1}{2}}^{3}f_{-\frac{1}{2}}-\\
        -2(Q/2+P)^{2}(Q+P)G_{-\frac{3}{2}}f_{-\frac{1}{2}}+
        2(Q/2+P+b)(Q/2+P+b^{-1})(Q+P)G_{-\frac{1}{2}}f_{-\frac{3}{2}}+\\+
        2(Q/2+P)(Q/2+P+b)(Q/2+P+b^{-1})(Q+P)f_{-\frac{1}{2}}f_{-\frac{3}{2}}\bigr)|P\rangle_{\scriptscriptstyle{\textsf{NS}}},
\end{multline}
\begin{multline}
        |P,-2\rangle=\bigl(G_{-\frac{1}{2}}^{4}+(Q/2-P)^{2}G_{-\frac{1}{2}}G_{-\frac{3}{2}}-
        (Q/2-P+b)(Q/2-P+b^{-1})G_{-\frac{3}{2}}G_{-\frac{1}{2}}
        -2(Q-P)G_{-\frac{1}{2}}^{3}f_{-\frac{1}{2}}-\\
        -2(Q/2-P)^{2}(Q-P)G_{-\frac{3}{2}}f_{-\frac{1}{2}}+
        2(Q/2-P+b)(Q/2-P+b^{-1})(Q-P)G_{-\frac{1}{2}}f_{-\frac{3}{2}}+\\+
        2(Q/2-P)(Q/2-P+b)(Q/2-P+b^{-1})(Q-P)f_{-\frac{1}{2}}f_{-\frac{3}{2}}\bigr)|P\rangle_{\scriptscriptstyle{\textsf{NS}}}.
\end{multline}
\end{subequations}
We note that there is an obvious relation
\begin{equation}\label{k-k}
  |P,k\rangle=|-P,-k\rangle.
\end{equation}
For general values of integer number $k$ one can construct the state $|P,k\rangle$ as described in section \ref{SAGT}. Due to \eqref{k-k} it is enough to consider only the case $k>0$. Then we can look for the expression for the vector $|P,k\rangle$ in the form
\begin{equation}\label{high-weight-def1-app}
  |P,k\rangle=(G_{-\frac{1}{2}}^{k^{2}}+C_{1}(P)G_{-\frac{1}{2}}^{k^{2}-3}G_{-\frac{3}{2}}+\dots)|P\rangle_{\scriptscriptstyle{\textsf{NS}}},
\end{equation}
where  $(C_{1}(P)\dots)$ are the coefficients to be determined. As was explained in section \ref{SAGT} the state $|P,k\rangle$ has nice representation in terms of free fields. That meansthet if we express generators $G_{r}$ as in \eqref{NS-bosonization} (for $k>0$ we have to take the sign ``$-$'' in  \eqref{NS-bosonization}) and use commutation relations \eqref{NS-bosonization-comm-relat} we will have
\begin{equation}\label{high-weight-def2-app}
   |P,k\rangle=\Omega_{k}(P)\,\chi_{-\frac{1}{2}}\chi_{-\frac{3}{2}}\dots\chi_{-\frac{2|k|-1}{2}}|\text{\textrm{vac}}\rangle,
\end{equation}
where
\begin{equation*}
   \chi_{r}=f_{r}-i\psi_{r}.
\end{equation*}
Comparing \eqref{high-weight-def2-app} and \eqref{high-weight-def1-app} we find all the coefficients $C_{j}(P)$ unambiguously.
\section{Comparing of $Z_{\textrm{pure}}^{X_{2}}$ and $Z_{\textrm{pure}}^{\diamond}$}\label{Bersh-app}
We claimed in section \ref{SAGT-collored} that the sets of summands on the left hand side and on the right hand side of the identity \eqref{eq Z=Z} are different. In this appendix we give an example of such phenomena. The expressions in \eqref{eq Z=Z}  differs first time in coefficient $\Lambda^{8}$ of $\Lambda$ expansion. For shortness we will use following notation:
$$\eps_{i,\,j}=i\eps_1+j\eps_2,\quad a_{i,\,j}=2a+i\eps_1+j\eps_2.$$
The left hand side of \eqref{eq Z=Z} can be computed using the formula \eqref{eq det2 color} (we omit $\vec{a}, \eps_1, \eps_2$ in notation). In the order $\Lambda^{8}$ the result reads:
\begin{align*}
&Z_{\textsf{vec}}^{\diamond}((4),\,\varnothing)+ Z_{\textsf{vec}}^{\diamond}((3,1),\,\varnothing)+ Z_{\textsf{vec}}^{\diamond}(((2,2),\,\varnothing)+ Z_{\textsf{vec}}^{\diamond}(((2,1,1),\,\varnothing)+\\ & Z_{\textsf{vec}}^{\diamond}(((1,1,1,1),\,\varnothing)+Z_{\textsf{vec}}^{\diamond}((2,1),\,(1))+ Z_{\textsf{vec}}^{\diamond}(((2),\,(2))+ Z_{\textsf{vec}}^{\diamond}(((2),\,(1,1))+\\ & Z_{\textsf{vec}}^{\diamond}((1,1),\,(2))+ Z_{\textsf{vec}}^{\diamond}((1,1),\,(1,1))+ Z_{\textsf{vec}}^{\diamond}(((1),\,(2,1))+ Z_{\textsf{vec}}^{\diamond}(\varnothing\,(4))+\\ &  Z_{\textsf{vec}}^{\diamond}(\varnothing,\,(3,1))+ Z_{\textsf{vec}}^{\diamond}(\varnothing,\,(2,2))+ Z_{\textsf{vec}}^{\diamond}(\varnothing,\,(2,1,1))+  Z_{\textsf{vec}}^{\diamond}(\varnothing,\,(1,1,1,1))=\\ &\frac1{\eps_{1,\,-3}\eps_{0,\,4}\eps_{1,\,-1}\eps_{0,\,2} a_{1,\,1}a_{0,\,0}a_{1,\,3}a_{0,\,2}}+
\frac1{\eps_{2,\,-2}\eps_{-1,\,3}\eps_{1,\,-1}\eps_{0,\,2} a_{1,\,3}a_{0,\,2}a_{1,\,1}a_{0,\,0}}+
\\& \frac1{\eps_{2,\,0}\eps_{-1,\,1}\eps_{1,\,-1}\eps_{0,\,2} a_{2,\,2}a_{1,\,1}a_{1,\,1}a_{0,\,0}}+
\frac1{\eps_{3,\,-1}\eps_{-2,\,2}\eps_{2,\,0}\eps_{-1,\,1} a_{3,\,1}a_{2,\,0}a_{1,\,1}a_{0,\,0}}+
\\& \frac1{\eps_{4,\,0}\eps_{-3,\,1}\eps_{2,\,0}\eps_{-1,\,1} a_{3,\,1}a_{2,\,0}a_{1,\,1}a_{0,\,0}}+
\frac1{a_{2,\,0}a_{1,\,-1}a_{1,\,1}a_{-0,\,0} a_{1,\,1}a_{0,\,0}a_{-1,\,1}a_{0,\,2}}+\\
&\frac1{\eps_{1,\,-1}\eps_{0,\,2}\eps_{1,\,-1}\eps_{0,\,2} a_{1,\,-1}a_{0,\,-2}a_{-1,\,1}a_{0,\,2}}+
\frac1{\eps_{1,\,-1}\eps_{0,\,2}\eps_{2,\,0}\eps_{-1,\,1} a_{1,\,1}a_{0,\,0}a_{1,\,1}a_{0,\,0}}+
\\& \frac1{\eps_{1,\,-1}\eps_{0,\,2}\eps_{2,\,0}\eps_{-1,\,1} a_{1,\,1}a_{0,\,0}a_{1,\,1}a_{0,\,0}}+
\frac1{\eps_{2,\,0}\eps_{-1,\,1}\eps_{2,\,0}\eps_{-1,\,1} a_{2,\,0}a_{1,\,-1}a_{-2,\,0}a_{-1,\,1}}+
\\& \frac1{a_{1,\,-1}a_{0,\,2}a_{-2,\,0}a_{-1,\,1} a_{-1,\,-1}a_{0,\,0}a_{-1,\,-1}a_{0,\,0}}+
\frac1{\eps_{1,\,-3}\eps_{0,\,4}\eps_{1,\,-1}\eps_{0,\,2} a_{-1,\,-1}a_{0,\,0}a_{-1,\,-3}a_{0,\,-2}}+\\
&\frac1{\eps_{2,\,-2}\eps_{-1,\,3}\eps_{1,\,-1}\eps_{0,\,2} a_{-1,\,-3}a_{0,\,-2}a_{-1,\,-1}a_{0,\,0}}+
\frac1{\eps_{2,\,0}\eps_{-1,\,1}\eps_{1,\,-1}\eps_{0,\,2} a_{-2,\,-2}a_{-1,\,-1}a_{-1,\,-1}a_{0,\,0}}+\\
& \frac1{\eps_{3,\,-1}\eps_{-2,\,2}\eps_{2,\,0}\eps_{-1,\,1} a_{-3,\,-1}a_{-2,\,0}a_{-1,\,-1}a_{0,\,0}}+
\frac1{\eps_{4,\,0}\eps_{-3,\,1}\eps_{2,\,0}\eps_{-1,\,1} a_{-3,\,-1}a_{-2,\,0}a_{-1,\,-1}a_{0,\,0}}=
\\& \frac{16 a^4-52 a^2 \epsilon _1^2+36 \epsilon _1^4-92 a^2 \epsilon _1 \epsilon _2+177 \epsilon _1^3 \epsilon _2-52 a^2 \epsilon _2^2+294 \epsilon _1^2 \epsilon _2^2+177 \epsilon _1 \epsilon _2^3+36 \epsilon _2^4}{2 \eps_1 \eps_2 a_{-1,\,-1} a_{1,\,1} a_{-2,\,-2} a_{2,\,2} a_{-3,\,-1} a_{-1,\,-3} a_{1,\,3} a_{3,\,1}}\;.
\end{align*}
The right hand side of \eqref{eq Z=Z} can be computed using the formula \eqref{Zvec-def-2}
\begin{align*}
&Z_{\textsf{vec}}(\{\varnothing,\,\varnothing\},\,\{\varnothing,\,\varnothing\},\,-2)+
Z_{\textsf{vec}}(\{(2),\,\varnothing\},\,\{\varnothing,\,\varnothing\},\,0)+ Z_{\textsf{vec}}(\{(1,1),\,\varnothing\},\,\{\varnothing,\,\varnothing\},\,0)+ \\ & Z_{\textsf{vec}}(\{\{1\},\,\{1\}\},\,\{\varnothing,\,\varnothing\},\,0)+ Z_{\textsf{vec}}(\{\varnothing,\,(2)\},\,\{\varnothing,\,\varnothing\},\,0)+ Z_{\textsf{vec}}(\{\varnothing,\,(1,1)\},\,\{\varnothing,\,\varnothing\},\,0)+ \\ & Z_{\textsf{vec}}(\{(1),\,\varnothing\},\,\{(1),\,\varnothing\},\,0)+ Z_{\textsf{vec}}(\{(1),\,\varnothing\},\,\{\varnothing,\,(1)\},\,0)+
Z_{\textsf{vec}}(\{\varnothing,\,(1)\},\,\{(1),\,\varnothing\},\,0)+ \\ & Z_{\textsf{vec}}(\{\varnothing,\,(1)\},\,\{\varnothing,\,(1)\},\,0)+ Z_{\textsf{vec}}(\{\varnothing,\,\varnothing\},\,\{(2),\,\varnothing\},\,0)+ Z_{\textsf{vec}}(\{\varnothing,\,\varnothing\},\,\{(1,1),\,\varnothing\},\,0)+ \\ & Z_{\textsf{vec}}(\{\varnothing,\,\varnothing\},\,\{(1),\,(1)\},\,0)+ Z_{\textsf{vec}}(\{\varnothing,\,\varnothing\},\,\{\varnothing,\,(2)\},\,0)+ Z_{\textsf{vec}}(\{\varnothing,\,\varnothing\},\,\{\varnothing,\,(1,1)\},\,0)+ \\ & Z_{\textsf{vec}}(\{\varnothing,\,\varnothing\},\,\{\varnothing,\,\varnothing\},\,2)=\\
&\frac1{a_{0,\,0}a_{-2,\,0}a_{0,\,-2} a_{-1,\,-1}a_{-1,\,-1} a_{-1,\,-3} a_{-3,\,-1}a_{-2,\,-2}}+
\frac1{\eps_{3,\,-1}\eps_{-2,\,2}\eps_{2,\,0}\eps_{-1,\,1} a_{1,\,1}a_{0,\,0}a_{0,\,2}a_{-1,\,1}}+\\
& \frac1{\eps_{4,\,0}\eps_{-3,\,1}\eps_{2,\,0}\eps_{-1,\,1} a_{3,\,1}a_{2,\,0}a_{1,\,1}a_{0,\,0}}+
\frac1{\eps_{2,\,0}\eps_{-1,\,1}\eps_{2,\,0}\eps_{-1,\,1} a_{2,\,0}a_{1,\,-1}a_{-2,\,0}a_{-1,\,1}}+\\
& \frac1{\eps_{3,\,-1}\eps_{-2,\,2}\eps_{2,\,0}\eps_{-1,\,1} a_{-1,\,-1}a_{0,\,0}a_{0,\,-2}a_{1,\,-1}}+
\frac1{\eps_{4,\,0}\eps_{-3,\,1}\eps_{2,\,0}\eps_{-1,\,1} a_{-3,\,-1}a_{-2,\,0}a_{-1,\,-1}a_{0,\,0}}+\\
& \frac1{\eps_{2,\,0}\eps_{-1,\,1}\eps_{1,\,-1}\eps_{0,\,2} a_{1,\,1}a_{0,\,0}a_{1,\,1}a_{0,\,0}}+
\frac1{\eps_{2,\,0}\eps_{-1,\,1}\eps_{1,\,-1}\eps_{0,\,2} a_{1,\,1}a_{0,\,0}a_{-1,\,-1}a_{0,\,0}}+\\
& \frac1{\eps_{-1,\,1}\eps_{0,\,2}\eps_{2,\,0}\eps_{-1,\,1} a_{1,\,1}a_{0,\,0}a_{-1,\,-1}a_{0,\,0}}+
\frac1{\eps_{2,\,0}\eps_{-1,\,1}\eps_{1,\,-1}\eps_{0,\,2} a_{-1,\,-1}a_{0,\,0}a_{-1,\,-1}a_{0,\,0}}+
\end{align*}
\begin{align*}
& \frac1{\eps_{1,\,-3}\eps_{0,\,4}\eps_{1,\,-1}\eps_{0,\,2} a_{1,\,1}a_{0,\,0}a_{1,\,3}a_{0,\,2}}+
\frac1{\eps_{2,\,-2}\eps_{-1,\,3}\eps_{1,\,-1}\eps_{0,\,2} a_{2,\,0}a_{1,\,-1}a_{1,\,1}a_{0,\,0}}+\\
& \frac1{\eps_{1,\,-1}\eps_{0,\,2}\eps_{1,\,-1}\eps_{0,\,2} a_{1,\,-1}a_{0,\,-2}a_{-1,\,1}a_{0,\,2}}+
\frac1{\eps_{1,\,-3}\eps_{0,\,4}\eps_{1,\,-1}\eps_{0,\,2} a_{-1,\,-1}a_{0,\,0}a_{-1,\,-3}a_{0,\,-2}}+\\
& \frac1{\eps_{2,\,-2}\eps_{-1,\,3}\eps_{1,\,-1}\eps_{0,\,2} a_{-2,\,0}a_{-1,\,1}a_{-1,\,-1}a_{0,\,0}}+
\frac1{a_{0,\,0}a_{2,\,0}a_{0,\,2} a_{1,\,1}a_{1,\,1} a_{1,\,3} a_{3,\,1}a_{2,\,2}}=\\
&\frac{16 a^4-52 a^2 \epsilon _1^2+36 \epsilon _1^4-92 a^2 \epsilon _1 \epsilon _2+177 \epsilon _1^3 \epsilon _2-52 a^2 \epsilon _2^2+294 \epsilon _1^2 \epsilon _2^2+177 \epsilon _1 \epsilon _2^3+36 \epsilon _2^4}{2 \eps_1 \eps_2 a_{-1,\,-1} a_{1,\,1} a_{-2,\,-2} a_{2,\,2} a_{-3,\,-1} a_{-1,\,-3} a_{1,\,3} a_{3,\,1}}\;.
\end{align*}
We see that results are the same but the sets of summands are different. For example there are only two summands which have degree $8$ in variable $a$, but these summands are different.
\bibliographystyle{MyStyle}
\bibliography{MyBib}

\providecommand{\href}[2]{#2}\begingroup\raggedright\begin{thebibliography}{10}

\bibitem{Alday:2009aq}
L.~F. Alday, D.~Gaiotto, and Y.~Tachikawa, {\it {Liouville correlation
  functions from four-dimensional gauge theories}},  {\em Lett. Math. Phys.}
  {\bf 91} (2010) 167--197, [\href{http://xxx.lanl.gov/abs/0906.3219}{{\tt
  arXiv:0906.3219}}].

\bibitem{Nekrasov:2002qd}
N.~A. Nekrasov, {\it {Seiberg-Witten Prepotential From Instanton Counting}},
  {\em Adv. Theor. Math. Phys.} {\bf 7} (2004) 831--864,
  [\href{http://xxx.lanl.gov/abs/hep-th/0206161}{{\tt hep-th/0206161}}].

\bibitem{Nakajima_1995}
H.~Nakajima, {\it {Heisenberg algebra and Hilbert schemes of points on
  projective surfaces}},  {\em Ann. of Math.} {\bf 145} (1997) 379--388,
  [\href{http://xxx.lanl.gov/abs/alg-geom/9507012}{{\tt alg-geom/9507012}}].

\bibitem{0970.17017}
H.~Nakajima, {\it {Quiver varieties and Kac-Moody algebras}},  {\em Duke Math.
  J.} {\bf 91} (1998) 515--560.

\bibitem{Atiyah_Bott_1984}
M.~Atiyah and R.~Bott, {\it The moment map and equivariant cohomology},  {\em
  Topology} {\bf 23} (1984) 1--28.

\bibitem{Belavin:2011pp}
V.~Belavin and B.~Feigin, {\it {Super Liouville conformal blocks from N=2 SU(2)
  quiver gauge theories}},  {\em JHEP} {\bf 1107} (2011) 079,
  [\href{http://xxx.lanl.gov/abs/1105.5800}{{\tt arXiv:1105.5800}}].

\bibitem{Goddard:1986ee}
P.~Goddard, A.~Kent, and D.~I. Olive, {\it {Unitary representations of the
  Virasoro and Supervirasoro algebras}},  {\em Commun. Math. Phys.} {\bf 103}
  (1986) 105--119.

\bibitem{Nakajima:fk}
H.~Nakajima, {\it Quiver varieties and finite dimensional representations of
  quantum affine algebras},  {\em J. Amer. Math. Soc.} {\bf 14} (2001)
  145--238, [\href{http://xxx.lanl.gov/abs/math/9912158}{{\tt math/9912158}}].

\bibitem{10.1063/1.2823979}
K.~Miki, {\it {$A(q,\gamma)$ analog of the $W_{1+\infty}$ algebra}},  {\em J.
  Math. Phys.} {\bf 48} (2007) 123520.

\bibitem{2010arXiv1002.2485F}
B.~{Feigin}, A.~{Hoshino}, J.~{Shibahara}, J.~{Shiraishi}, and S.~{Yanagida},
  {\it {Kernel function and quantum algebras}},  {\em RIMS Kokyuroku} {\bf
  1689} (2010) 133--152, [\href{http://xxx.lanl.gov/abs/1002.2485}{{\tt
  arXiv:1002.2485}}].

\bibitem{Feigin-unpub}
B.~Feigin \hspace*{-3pt}, unpublished.

\bibitem{Awata:2011fk}
H.~Awata, B.~Feigin, A.~Hoshino, M.~Kanai, J.~Shiraishi, and S.~Yanagida, {\it
  {Notes on Ding-Iohara algebra and AGT conjecture}},
  \href{http://xxx.lanl.gov/abs/1106.4088}{{\tt arXiv:1106.4088}}.

\bibitem{Li:uq}
W.-P. Li, Z.~Qin, and W.~Wang, {\it {The cohomology rings of Hilbert schemes
  via Jack polynomials}},  \href{http://xxx.lanl.gov/abs/math/0411255v1}{{\tt
  math/0411255v1}}.

\bibitem{Carlsson:2008fk}
E.~Carlsson and A.~Okounkov, {\it Exts and vertex operators},
  \href{http://xxx.lanl.gov/abs/0801.2565}{{\tt arXiv:0801.2565}}.

\bibitem{Alday:2010vg}
L.~F. Alday and Y.~Tachikawa, {\it {Affine SL(2) conformal blocks from 4d gauge
  theories}},  {\em Lett. Math. Phys.} {\bf 94} (2010) 87--114,
  [\href{http://xxx.lanl.gov/abs/1005.4469}{{\tt arXiv:1005.4469}}].

\bibitem{Alba:2010qc}
V.~A. Alba, V.~A. Fateev, A.~V. Litvinov, and G.~M. Tarnopolsky, {\it {On
  combinatorial expansion of the conformal blocks arising from AGT
  conjecture}},  {\em Lett. Math. Phys.} {\bf 98} (2011) 33--64,
  [\href{http://xxx.lanl.gov/abs/1012.1312}{{\tt arXiv:1012.1312}}].

\bibitem{Wyllard:2009hg}
N.~Wyllard, {\it {$A_{N-1}$ conformal Toda field theory correlation functions
  from conformal $N=2$ $SU(N)$ quiver gauge theories}},  {\em JHEP} {\bf 11}
  (2009) 002, [\href{http://xxx.lanl.gov/abs/0907.2189}{{\tt
  arXiv:0907.2189}}].

\bibitem{Mironov:2009by}
A.~Mironov and A.~Morozov, {\it {On AGT relation in the case of $U(3)$}},  {\em
  Nucl. Phys.} {\bf B825} (2010) 1--37,
  [\href{http://xxx.lanl.gov/abs/0908.2569}{{\tt arXiv:0908.2569}}].

\bibitem{Fateev:2011hq}
V.~A. Fateev and A.~V. Litvinov, {\it {Integrable structure, W-symmetry and AGT
  relation}},  {\em JHEP} {\bf 01} (2012) 051,
  [\href{http://xxx.lanl.gov/abs/1109.4042}{{\tt arXiv:1109.4042}}].

\bibitem{Belavin:2011tb}
A.~Belavin, V.~Belavin, and M.~Bershtein, {\it {Instantons and 2d
  Superconformal field theory}},  {\em JHEP} {\bf 1109} (2011) 117,
  [\href{http://xxx.lanl.gov/abs/1106.4001}{{\tt arXiv:1106.4001}}].

\bibitem{Ito:2011mw}
Y.~Ito, {\it {Ramond sector of super Liouville theory from instantons on an ALE
  space}},  {\em Nucl.Phys.} {\bf B861} (2012) 387--402,
  [\href{http://xxx.lanl.gov/abs/1110.2176}{{\tt arXiv:1110.2176}}].

\bibitem{Nishioka:2011jk}
T.~Nishioka and Y.~Tachikawa, {\it {Para-Liouville/Toda central charges from
  M5-branes}},  {\em Phys. Rev.} {\bf D84} (2011) 046009,
  [\href{http://xxx.lanl.gov/abs/1106.1172}{{\tt arXiv:1106.1172}}].

\bibitem{Wyllard:2011mn}
N.~Wyllard, {\it {Coset conformal blocks and $N=2$ gauge theories}},
  \href{http://xxx.lanl.gov/abs/1109.4264}{{\tt arXiv:1109.4264}}.

\bibitem{Alfimov:2011ju}
M.~Alfimov and G.~Tarnopolsky, {\it {Parafermionic Liouville field theory and
  instantons on ALE spaces}},  {\em JHEP} {\bf 1202} (2012) 036,
  [\href{http://xxx.lanl.gov/abs/1110.5628}{{\tt arXiv:1110.5628}}].

\bibitem{Bonelli:2011jx}
G.~Bonelli, K.~Maruyoshi, and A.~Tanzini, {\it {Instantons on ALE spaces and
  Super Liouville Conformal Field Theories}},  {\em JHEP} {\bf 1108} (2011)
  056, [\href{http://xxx.lanl.gov/abs/1106.2505}{{\tt arXiv:1106.2505}}].

\bibitem{Bonelli:2011kv}
G.~Bonelli, K.~Maruyoshi, and A.~Tanzini, {\it {Gauge Theories on ALE Space and
  Super Liouville Correlation Functions}},
  \href{http://xxx.lanl.gov/abs/1107.4609}{{\tt arXiv:1107.4609}}.

\bibitem{Argyres:1990aq}
P.~C. Argyres, A.~LeClair, and S.~H.~H. Tye, {\it {On the possibility of
  fractional superstrings}},  {\em Phys. Lett.} {\bf B253} (1991) 306--312.

\bibitem{Fateev:1985ig}
V.~A. Fateev and A.~B. Zamolodchikov, {\it {Representations of the algebra of
  parafermion currents of spin $4/3$ in two-dimensional conformal field theory.
  Minimal models and the tricritical Potts $Z(3)$ model}},  {\em Theor. Math.
  Phys.} {\bf 71} (1987) 451--462.

\bibitem{Pogosian:1988ar}
R.~G. Pogosian, {\it {Operator algebra in two-dimensional conformal quantum
  field theory containing spin $4/3$ parafermionic conserved currents.}},  {\em
  Int. J. Mod. Phys.} {\bf A6} (1991) 2005--2023.

\bibitem{0949.14001}
H.~Nakajima, {\em {Lectures on Hilbert schemes of points on surfaces.}}
\newblock {University Lecture Series. 18. Providence, RI: American Mathematical
  Society (AMS). xi, 132 p.}, 1999.

\bibitem{2003math.....11058N}
H.~{Nakajima} and K.~{Yoshioka}, {\it {Lectures on Instanton Counting}},
  \href{http://xxx.lanl.gov/abs/math/0311058}{{\tt math/0311058}}.

\bibitem{Flume:2002az}
R.~Flume and R.~Poghossian, {\it {An algorithm for the microscopic evaluation
  of the coefficients of the Seiberg-Witten prepotential}},  {\em Int. J. Mod.
  Phys.} {\bf A18} (2003) 2541,
  [\href{http://xxx.lanl.gov/abs/hep-th/0208176}{{\tt hep-th/0208176}}].

\bibitem{2003math......6198N}
H.~{Nakajima} and K.~{Yoshioka}, {\it {Instanton counting on blowup. I.
  4-dimensional pure gauge theory}},  {\em Invent. Math.} {\bf 162} (2005)
  313--355, [\href{http://xxx.lanl.gov/abs/math/0306198}{{\tt math/0306198}}].

\bibitem{Fucito:2004gi}
F.~Fucito, J.~F. Morales, and R.~Poghossian, {\it {Instantons on quivers and
  orientifolds}},  {\em JHEP} {\bf 10} (2004) 037,
  [\href{http://xxx.lanl.gov/abs/hep-th/0408090}{{\tt hep-th/0408090}}].

\bibitem{Shadchin:2005cc}
S.~Shadchin, {\it {Cubic curves from instanton counting}},  {\em JHEP} {\bf 03}
  (2006) 046, [\href{http://xxx.lanl.gov/abs/hep-th/0511132}{{\tt
  hep-th/0511132}}].

\bibitem{Macdonald}
I.~G. Macdonald, {\em {Symmetric functions and Hall polynomials}}.
\newblock Oxford University Press, 1995.

\bibitem{Belavin:2011js}
A.~Belavin and V.~Belavin, {\it {AGT conjecture and Integrable structure of
  Conformal field theory for c=1}},  {\em Nucl. Phys.} {\bf B850} (2011)
  199--213, [\href{http://xxx.lanl.gov/abs/1102.0343}{{\tt arXiv:1102.0343}}].

\bibitem{Estienne:2011qk}
B.~Estienne, V.~Pasquier, R.~Santachiara, and D.~Serban, {\it {Conformal blocks
  in Virasoro and W theories: Duality and the Calogero-Sutherland model}},
  {\em Nucl.Phys.} {\bf B860} (2012) 377--420,
  [\href{http://xxx.lanl.gov/abs/1110.1101}{{\tt arXiv:1110.1101}}].

\bibitem{Zamolodchikov:1995aa}
A.~B. Zamolodchikov and {\relax Al}.~B. Zamolodchikov, {\it Structure constants
  and conformal bootstrap in {L}iouville field theory},  {\em Nucl. Phys.} {\bf
  B477} (1996) 577--605, [\href{http://xxx.lanl.gov/abs/hep-th/9506136}{{\tt
  hep-th/9506136}}].

\bibitem{1166.14007}
H.~Nakajima, {\it {Sheaves on ALE spaces and quiver varieties}},  {\em Mosc.
  Math. J.} {\bf 7} (2007) 699--722.

\bibitem{2011CMaPh.304..395B}
U.~{Bruzzo}, R.~{Poghossian}, and A.~{Tanzini}, {\it {Poincar{\'e} Polynomial
  of Moduli Spaces of Framed Sheaves on (Stacky) Hirzebruch Surfaces}},  {\em
  Commun. Math. Phys.} {\bf 304} (2011) 395--409,
  [\href{http://xxx.lanl.gov/abs/0909.1458}{{\tt arXiv:0909.1458}}].

\bibitem{Crnkovic:1989gy}
C.~Crnkovic, G.~Sotkov, and M.~Stanishkov, {\it {Renormalization group flow for
  general $SU(2)$ coset models}},  {\em Phys.Lett.} {\bf B226} (1989) 297.

\bibitem{Crnkovic:1989ug}
C.~Crnkovic, R.~Paunov, G.~Sotkov, and M.~Stanishkov, {\it {Fusions of
  conformal models}},  {\em Nucl.Phys.} {\bf B336} (1990) 637.

\bibitem{Lashkevich:1992sb}
M.~Lashkevich, {\it {Superconformal 2-D minimal models and an unusual coset
  construction}},  {\em Mod. Phys. Lett.} {\bf A8} (1993) 851--860,
  [\href{http://xxx.lanl.gov/abs/hep-th/9301093}{{\tt hep-th/9301093}}].

\bibitem{Fucito:2004ry}
F.~Fucito, J.~F. Morales, and R.~Poghossian, {\it {Multi instanton calculus on
  ALE spaces}},  {\em Nucl. Phys.} {\bf B703} (2004) 518--536,
  [\href{http://xxx.lanl.gov/abs/hep-th/0406243}{{\tt hep-th/0406243}}].

\bibitem{Fucito:2006kn}
F.~Fucito, J.~F. Morales, and R.~Poghossian, {\it {Instanton on toric
  singularities and black hole countings}},  {\em JHEP} {\bf 12} (2006) 073,
  [\href{http://xxx.lanl.gov/abs/hep-th/0610154}{{\tt hep-th/0610154}}].

\bibitem{Nagao_2007}
K.~Nagao, {\it {Quiver varieties and Frenkel-Kac construction}},  {\em Journal
  of Algebra} {\bf 321} (2007) 3764--3789,
  [\href{http://xxx.lanl.gov/abs/math/0703107}{{\tt math/0703107}}].

\bibitem{Pog-cite}
R.~Poghossian \hspace*{-3pt}, unpublished.

\bibitem{Zamolodchikov:2005fy}
{\relax Al}.~B. Zamolodchikov, {\it {Three-point function in the minimal
  Liouville gravity}},  {\em Theor. Math. Phys.} {\bf 142} (2005) 183--196,
  [\href{http://xxx.lanl.gov/abs/hep-th/0505063}{{\tt hep-th/0505063}}].

\bibitem{Rashkov:1996jx}
R.~C. Rashkov and M.~Stanishkov, {\it {Three-point correlation functions in
  $N=1$ Super Lioville Theory}},  {\em Phys. Lett.} {\bf B380} (1996) 49--58,
  [\href{http://xxx.lanl.gov/abs/hep-th/9602148}{{\tt hep-th/9602148}}].

\bibitem{Poghosian:1996dw}
R.~H. Poghosian, {\it {Structure constants in the N = 1 super-Liouville field
  theory}},  {\em Nucl. Phys.} {\bf B496} (1997) 451--464,
  [\href{http://xxx.lanl.gov/abs/hep-th/9607120}{{\tt hep-th/9607120}}].

\bibitem{Bershtein:2010wz}
M.~A. Bershtein, V.~A. Fateev, and A.~V. Litvinov, {\it {Parafermionic
  polynomials, Selberg integrals and three- point correlation function in
  parafermionic Liouville field theory}},  {\em Nucl. Phys.} {\bf B847} (2011)
  413--459, [\href{http://xxx.lanl.gov/abs/1011.4090}{{\tt arXiv:1011.4090}}].

\bibitem{Fateev:1996ea}
V.~A. Fateev, {\it {The sigma model (dual) representation for a two-parameter
  family of integrable quantum field theories}},  {\em Nucl. Phys.} {\bf B473}
  (1996) 509--538.

\end{thebibliography}\endgroup
\end{document}